\documentclass[journal]{IEEEtran}

\usepackage{cite,graphicx,amsmath,amssymb}
\usepackage{color,psfrag}

\usepackage{citesort}
\usepackage{comment}
\usepackage{cases}
\usepackage{accents}
\usepackage{stfloats}
\usepackage{times,relsize}
\usepackage{latexsym}

\newtheorem{theorem}{Theorem}
\newtheorem{lemma}{Lemma}
\newtheorem{corollary}{Corollary}

\newcommand{\bs}{\boldsymbol}
\newcommand{\bvap}{\boldsymbol{\Xi}}
\newcommand{\bxi}{\boldsymbol{\xi}}
\newcommand{\bOm}{\boldsymbol{\Omega}}
\newcommand{\bbe}{\mathbb{E}}
\newcommand{\bh}{\mathbf{h}}

\newcommand{\bH}{\mathbf{H}}

\newcommand{\bG}{\mathbf{G}}
\newcommand{\bg}{\mathbf{g}}
\newcommand{\bhG}{\mathbf{\hat G}}
\newcommand{\bhg}{\mathbf{\hat g}}

\newcommand{\ntsp}{\negthickspace}
\newcommand{\nmsp}{\negmedspace}

\newcommand{\asskip}{\abovedisplayshortskip}
\newcommand{\bsskip}{\belowdisplayshortskip}
\newcommand{\alskip}{\abovedisplayskip}
\newcommand{\blskip}{\belowdisplayskip}

\newcommand*{\dt}[1]{\accentset{\mbox{\large\bfseries .}}{#1}}

\begin{document}
\title{Power Scaling of Uplink Massive MIMO Systems with Arbitrary-Rank Channel Means}
\author{Qi~Zhang,
        Shi~Jin,~\IEEEmembership{Member,~IEEE,}
        Kai-Kit~Wong,~\IEEEmembership{Senior Member,~IEEE,} \\
        Hongbo~Zhu,
        and~Michail~Matthaiou,~\IEEEmembership{Senior Member,~IEEE}
\thanks{
        Manuscript received September 29, 2013; revised April 8, 2014.
        This work was partly supported by the China 973 project under Grant 2013CB329005, the National Natural Science Foundation of China under Grant 6127123 and
        the China 863 Program under Grant 2014AA01A705. The work of S. Jin was supported by the National Natural Science Foundation of China under Grant 61222102 and the Natural Science Foundation of Jiangsu Province under Grant BK2012021.
        This paper was presented in part at the IEEE
        Globe Communication Conference, Atlanta, USA, December 2013.}
\thanks{Copyright (c) 2014 IEEE. Personal use of this material is permitted. However, permission to use this material for any other purposes must be obtained from the IEEE by sending a request to pubs-permissions@ieee.org.}
\thanks{Q. Zhang and H. Zhu are with Jiangsu Key Laboratory of Wireless Communications, Nanjing University of Posts and Telecommunications, Nanjing, 210003, P. R. China (email: zhangqiqi$\_$1212@126.com; zhuhb@njupt.edu.cn).}
\thanks{S. Jin is with the National Mobile Communications Research Laboratory,
Southeast University, Nanjing, 210096, P. R. China (email: jinshi@seu.edu.cn). S. Jin is the corresponding author.}
\thanks{K.-K. Wong is with the Department of Electronic and Electrical Engineering, University College London,
London, WC1E 7JE, United Kingdom (email: kai-kit.wong@ucl.ac.uk).}
\thanks{M. Matthaiou is with the School of Electronics, Electrical Engineering and Computer Science, Queen¡¯s University Belfast, Belfast, BT3 9DT, U.K., and with the Department of Signals and Systems, Chalmers University of Technology, 412 96, Gothenburg, Sweden (e-mail: m.matthaiou@qub.ac.uk).}  }

\markboth{IEEE JOURNAL OF SELECTED TOPICS IN SIGNAL PROCESSING,~Vol.~x,
No.~xx,~xx~201x} {IEEE JOURNAL OF SELECTED TOPICS IN SIGNAL PROCESSING,~Vol.~x,
No.~xx,~xx~201x}

\maketitle

\asskip=3pt
\bsskip=3pt
\alskip=4.3pt
\blskip=4.3pt

\begin{abstract}
  This paper investigates the uplink achievable rates of massive multiple-input multiple-output (MIMO) antenna systems in Ricean fading channels, using maximal-ratio combining (MRC) and zero-forcing (ZF) receivers, assuming perfect and imperfect channel state information (CSI). In contrast to previous relevant works, the fast fading MIMO channel matrix is assumed to have an arbitrary-rank deterministic component as well as a Rayleigh-distributed random component. We derive tractable expressions for the achievable uplink rate in the large-antenna limit, along with approximating results that hold for any finite number of antennas. Based on these analytical results, we obtain the scaling law that the users' transmit power should satisfy, while maintaining a desirable quality of service. In particular, it is found that regardless of the Ricean $K$-factor, in the case of perfect CSI, the approximations converge to the same constant value as the exact results, as the number of base station antennas, $M$, grows large, while the transmit power of each user can be scaled down proportionally to $1/M$. If CSI is estimated with uncertainty, the same result holds true but only when the Ricean $K$-factor is non-zero. Otherwise, if the channel experiences Rayleigh fading, we can only cut the transmit power of each user proportionally to $1/\sqrt M$. In addition, we show that with an increasing Ricean $K$-factor, the uplink rates will converge to fixed values for both MRC and ZF receivers.
\end{abstract}


\section{Introduction}\label{sec:Intro}
Multiple-input multiple-output (MIMO) antenna technology has emerged as an effective technique for significantly improving the capacity of wireless communication systems \cite{paulraj03,tse05}. Recently, multiuser MIMO (MU-MIMO) systems, where a base station (BS) equipped with multiple antennas serves a number of users in the same time-frequency resource, have gained much attention because of their considerable spatial multiplexing gains even without multiple antennas at the users \cite{Gesbert07,viswanath03,caire10,jose11}. 
To reap all the benefits of MIMO at a greater scale, the paradigm of massive MIMO, which considers the use of hundreds of antenna elements to serve tens of users simultaneously, has recently come at the forefront of wireless communications research \cite{rusek13}.

 Great efforts have been made to understand the spectral and energy efficiency gains of massive MIMO systems, e.g., \cite{telatar99,marzetta06,marzetta10,Hoydis13,Ngo11,rusek13,pitarokoilis12,wagner10}. In particular, \cite{telatar99} indicates that the high number of degrees-of-freedom can average out the effects of fast fading. It has further been revealed in \cite{marzetta10} that, when the number of antennas increases without bound, uncorrelated noise, fast fading and intracell interference vanish. The only impairment left is pilot-contamination. Another merit of massive MIMO is that the transmit power can be greatly reduced.
In \cite{Ngo11}, the power-scaling law was investigated and it was shown that, as the number of BS antennas grows without limit, the uplink rate can be maintained while the transmit power can be substantially cut down. For example, ideally, to maintain the same quality-of-service as with a single-antenna BS, the transmit power of a 100-antenna BS would be only almost $1 \%$ of the power of the single-antenna one.

{ The increasing physical size of massive MIMO arrays is a fundamental problem for practical deployment and utilization and the millimeter-wave operating from $30-300$ GHz finds a way out, since the small wavelengths make possible for many antenna elements to be packed with a finite volume. On top of this, due to the highly directional and quasi-optical nature of propagation at millimeter-waves, line-of-sight (LOS) propagation is dominating \cite{sayeed11,brady13,sayeed13}.
 Therefore, massive MIMO systems operating in LOS conditions is expected to be a novel paradigm.} Unfortunately, many of the existing pioneering works simply assume Rayleigh fading conditions \cite{telatar99,marzetta10,Ngo11}.
While this assumption simplifies extensively all mathematical manipulations,
it falls short of capturing the fading variations when there is a specular or LOS component between the transmitter and receiver. As such, more general fading models need to be considered.

In this paper, 
we extend the results in \cite{Ngo11} to Ricean fading channels with arbitrary-rank mean matrices. In \cite{Ngo11}, the authors studied the potential of power savings in massive MU-MIMO systems, assuming that the fast fading channel matrix has zero-mean unit-variance entries. In our analysis, the fast fading channel matrix consists of an arbitrary-rank deterministic component, and a Rayleigh-distributed random component accounting for the scattered signals \cite{lozano03}. We consider a single-cell MU-MIMO system in the uplink, where both maximal-ratio combining (MRC)
and zero-forcing (ZF) receivers are assumed at the BS, with perfect and imperfect channel state information (CSI). { Some relevant works on Ricean fading in massive MIMO systems are \cite{jin07,zhang13,wen11,wen10}. However, \cite{jin07} considers the single-user scenario and
in \cite{zhang13,wen11,wen10}, the numbers of antennas at the transmitters and the receiver go to infinity
with a constant ratio. Our main contributions include new, tractable expressions for the achievable uplink rate in the large-antenna limit, along with approximating results that hold {\it for any finite number of antennas}.} We also elaborate on the power-scaling laws as follows:
\begin{itemize}
\item We reveal that under Ricean fading, with perfect CSI, if the number of BS antennas, $M$, grows asymptotically large, we can cut down the transmit power of each user proportionally to $1/M$ to maintain a desirable rate. In addition, as $M \to \infty$, the sum rates of both MRC and ZF receivers converge to the same constant value, indicating that in the large-system limit, intracell interference disappears.
\item If CSI is estimated with uncertainty, then when $M$ gets asymptotically large, massive MIMO will still bring considerate power savings for each user. In particular, if the Ricean $K$-factor is non-zero, the transmit power for each user can be scaled down by $1/M$ to obtain the same rate, while the uplink rates will tend to a fixed value as a function of Ricean $K$-factor. However, when the Ricean $K$-factor is zero, the transmit power can be scaled down only by $1/{\sqrt M}$ and the uplink rates will again approach to a fixed value if $M\to\infty$.
\end{itemize}


The remainder of the paper is organized as follows. {Section \ref{sec:model}} describes the MU-MIMO system model in Ricean fading channels, and provides the definition of the uplink rate with perfect and imperfect CSI. Section \ref{sec:achievable uplink rate} derives closed-form approximations for the achievable uplink rates and also investigates the power-scaling laws. In {Section \ref{sec:numerical results}}, we provide a set of numerical results, while Section \ref{sec:conclusion} summarizes the main results of this paper.

{\em Notation}---Throughout the paper, vectors are expressed in lowercase boldface letters while matrices are denoted by uppercase boldface letters. We use ${{\bf{X}}^H}, {{\bf{X}}^T}, {\bf{X}^*}$ and $\mathbf{X}^{-1}$ to denote the conjugate-transpose, transpose, conjugate and inverse of $\bf{X}$, respectively. Moreover, $\mathbf{I}_N$ denotes an $N\times N$ identity matrix, $\delta_{ni}$ equals $1$ when $n=i$ and $0$ otherwise, and $[{\bf{X}}]_{ij}$ or ${\bf{X}}_{ij}$ gives the ($i,j$)th entry of $\bf{X}$. Finally, $\mathbb{E}\left\{  \cdot  \right\}$ is the expectation operator, $\left\| {\, \cdot \,} \right\|$ is the Euclidean norm and $\mathbf{Z} \sim {\cal CN}\left(\mathbf{A},\mathbf{B}\right)$ denotes that $\mathbf{Z}$ is a complex Gaussian matrix with mean matrix $\mathbf{A}$ and covariance matrix $\mathbf{B}$.

\section{System Model}\label{sec:model}
We consider a MU-MIMO system with $N$ single-antenna users and an $M$-antenna BS, where users transmit their signals to the BS in the same time-frequency channel. The system is single-cell with no interference from neighboring cells. The received vector ${\mathbf{y}} \in {{\mathbb{C}}^{M\times 1}}$ at the BS can be written as \cite{Hoydis13}
\asskip=-4pt
\begin{equation}\label{system model}
{\mathbf{y}} = \sqrt {{p_u}} {\mathbf{Gx}} + {\mathbf{n}},
\end{equation}
where $\mathbf{G}$ denotes the $M \times N$ MIMO channel matrix between the BS and the $N$ users, $\sqrt{p_u}\mathbf{x}$ denotes the $N \times 1$ vector containing the transmitted signals from all users, $p_u$ is the average transmitted power of each user, and $\mathbf{n}$ represents the vector of zero-mean additive white Gaussian noise (AWGN). To facilitate our analysis and without loss of generality, the noise variance is assumed to be $1$.

\subsection{Channel Model}\label{sec:channel model}
We denote the channel coefficient between {the $n$th user and the $m$th antenna of the BS} as ${g_{mn}} = {[\mathbf{G}]_{mn}}$, which embraces independent fast fading, geometric attenuation and log-normal shadow fading \cite{marzetta10} and can be expressed as
\begin{equation}
{g_{mn}} = {h_{mn}}\sqrt {{\beta _n}},
\end{equation}
where $h_{mn}$ is the fast fading element from the $n$th user to the $m$th antenna of the BS, while $\beta_n$ is the large-scale fading coefficient to model both the geometric attenuation and shadow fading, which is assumed to be constant across the antenna array. Under this model, we can write
\begin{equation}\label{channel matrix}
{\mathbf{G}} = {\mathbf{H}}{{\mathbf{D}}^{1/2}},
\end{equation}
where $\mathbf{H}$ denotes the $M \times N$ channel matrix modeling fast fading between the users and the BS, i.e., ${[\mathbf{H}]_{mn}} = {h_{mn}}$ and $\mathbf{D}$ is the $N \times N$ diagonal matrix with ${[\mathbf{D}]_{nn}} = {\beta_{n}}$. The fast fading matrix consists of two parts, namely a deterministic component corresponding to the LOS signal and a Rayleigh-distributed random component which accounts for the scattered signals. Moreover, the Ricean factor, represents the ratio of the power of the deterministic component to the power of the scattered components. Here, we assume that the Ricean $K$-factor of each user is different and the $n$th user's $K$-factor is denoted by $K_n$. Then, the fast fading matrix $\mathbf{H}$ can be written as \cite{Proakis01}
\begin{equation}\label{fast fading matrix}
{\mathbf{H}} = {\mathbf{\bar H}}{\left[ {\bOm {{\left( {\bOm +\mathbf{I}_N } \right)}^{ - 1}}} \right]^{1/2}} + {{\mathbf{H}}_w}{\left[ {{{\left( {\bOm + \mathbf{I}_N } \right)}^{ - 1}}} \right]^{1/2}},
\end{equation}
where $\bOm$ is a $N \times N$ diagonal matrix with $\left[\bOm\right]_{nn}=K_n$, ${\mathbf{H}_w}$ denotes the random component, the entries of which are independent and identically distributed (i.i.d.) Gaussian random variables with zero-mean, independent real and imaginary parts, each with variance $1/2$, and $\mathbf{\bar H}$ denotes the deterministic component, which was usually assumed to be rank-$1$ in previous studies, e.g., \cite{lozano03,alfano04mutual,matthaiou11,jin07}. In this paper, due to the assumption of multiple geographically distributed users, this constraint is relaxed and we let $\mathbf{\bar H}$ have an arbitrary rank as \cite{ravindran07}\footnote{Note that with the change of $\theta_n$, \eqref{Hbar entries model} can become arbitrary-rank.}
\begin{equation}\label{Hbar entries model}
{\left[ {\mathbf{\bar H}} \right]_{mn}} = {e^{-j(m - 1)\frac{{2\pi d}}{\lambda }{\sin\left(\theta _n\right)}}},
\end{equation}
where $d$ is the antenna spacing, $\lambda$ is the wavelength, and $\theta_i$ is the arrival angle of the $i$th user. For convenience, we will set $d=\lambda/2$ in the rest of this paper.\footnote{Since the physical size of the antenna array depends on the operating frequency, it can be very small even with large $M$ at high frequencies (e.g. $60$GHz communications \cite{torkildson11}).}

\subsection{Achievable Uplink Rate}
\subsubsection{Perfect CSI}
We first consider the case that the BS has perfect CSI. Let $\mathbf{A}$ be the $M \times N$ linear receiver matrix which depends on the channel matrix $\mathbf{G}$. The BS processes its received signal vector by multiplying it with the conjugate-transpose of the linear receiver as \cite{marzetta10}
\begin{equation}\label{using receiver}
{\mathbf{r}} = \mathbf{A}^H\mathbf{y}.
\end{equation}
Then, substituting (\ref{system model}) into \eqref{using receiver} gives
\begin{equation}
{\mathbf{r}} = \sqrt {{p_u}} {\mathbf{A}^H}\mathbf{Gx} + {\mathbf{A}^H}\mathbf{n}.
\end{equation}
The $n$th element of $\mathbf{r}$ can be written as
\begin{equation}
{r_n} = \sqrt {{p_u}} \mathbf{a}_n^H\mathbf{Gx} + \mathbf{a}_n^H\mathbf{n},
\end{equation}
where $\mathbf{a}_n$ is the $n$th column of $\mathbf{A}$. By the law of matrix multiplication, we further get
\begin{equation}
{r_n} = \sqrt {{p_u}} \mathbf{a}_n^H{\mathbf{g}_n}{x_n} + \sqrt {{p_u}} \sum_{i = 1\atop i \ne n}^N{\mathbf{a}_n^H{\mathbf{g}_i}{x_i}}  + \mathbf{a}_n^H\mathbf{n},
\end{equation}
where $x_n$ denotes the $n$th element of $\mathbf{x}$ and $\mathbf{g}_n$ is the $n$th column of $\mathbf{G}$.
Assuming an ergodic channel, the achievable uplink rate of the $n$th user is given by \cite{Ngo11}\footnote{Hereafter, the subscript ``{\sf P}'' will correspond to the case of perfect CSI, while the subscript ``{\sf IP}'' to the imperfect CSI case.}
\begin{equation}\label{perfect uplink rate}
 {R_{{\sf P},n}} = \bbe\left\{ {{{\log }_2}\left( {1 + \frac{{{p_u}{{\left| {\mathbf{a}_n^H{\mathbf{g}_n}} \right|}^2}}}{{{p_u}\sum_{i = 1\atop i \ne n}^N {{{\left| {\mathbf{a}_n^H{\mathbf{g}_i}} \right|}^2}}  + {{\left\| {{\mathbf{a}_n}} \right\|}^2}}}} \right)} \right\}.
\end{equation}
Therefore, the uplink sum rate per cell, measured in bits/s/Hz, can be defined as
\begin{equation}
C_{\sf P}=\sum\limits_{n = 1}^N {{R_{\sf {P},n}}}.
\end{equation}

\subsubsection{Imperfect CSI}\label{sec:imperfect CSI model}
In real situations, the channel matrix $\mathbf{G}$ is estimated at the BS. For the considered Ricean fading channel model, we assume that both the deterministic LOS component and the Ricean $K$-factor matrix $\bOm$ are perfectly known at both the transmitter and receiver,\footnote{{Since the LOS channels are hardly changing, the BS may estimate the LOS components during the previous transmission from the users to the BS. While the estimation of the Ricean $K$ factor in massive MIMO systems is an interesting topic for additional research, we do not pursue it herein due to space constraints.}} such that the estimate of $\mathbf{G}$ can be expressed as
\begin{equation}\label{Imperfect CSI model}
{\bf{\hat G}} = {\bf{\bar G}}{\left[ {\bOm {{\left( {\bOm +\mathbf{I}_N } \right)}^{ - 1}}} \right]^{1/2}} + {{\bf{\hat G}}_w}{\left[ {{{\left( {\bOm + \mathbf{I}_N } \right)}^{ - 1}}} \right]^{1/2}},
\end{equation}
where ${\bf{\bar G}}$ denotes the deterministic component of $\mathbf{G}$, i.e., ${\bf{\bar G}} \triangleq {\bf{\bar H}}{{\bf{D}}^{1/2}}$ and ${{\bf{\hat G}}_w}$ represents the estimate of the random component $\bG_w \triangleq \bH_w \mathbf{D}^{1/2}$. In this paper, we consider that the channel is estimated using uplink
pilots. Let an interval of length $\tau$ symbols be used for uplink training, where $\tau$ is smaller than the coherence time of the channel. In the training stage, all users simultaneously transmit orthogonal pilot
sequences of $\tau$ symbols, which can be stacked into a $\tau \times N$ matrix $\sqrt p_p \boldsymbol{\Phi}$ $\left(\tau \ge N\right)$, which satisfies
$\mathbf{F}^H \mathbf{F}=\mathbf{I}_N$, where $\mathbf{F} \triangleq \bs{\Phi}\left[\left(\bOm+\mathbf{I}_N \right)^{-1}\right]^{1/2}$ and $p_p=\tau p_u$ is the transmit pilot power. As a result, the BS receives the $M \times \tau$ noisy pilot matrix as
\begin{equation}\label{received pilot}
{\mathbf{Y}_p} = \sqrt {{p_p}} \mathbf{G}{\boldsymbol{\Phi} ^T} + \mathbf{N},
\end{equation}
where $\mathbf{N}$ represents the $M \times \tau$ AWGN matrix with i.i.d.~zero-mean and unit-variance elements. With \eqref{fast fading matrix}, we can remove the LOS component, which is assumed to be already known from \eqref{received pilot}, and the remaining terms of the received matrix are
\begin{equation}\label{rest received pilot}
\mathbf{Y}_{p,w}= \sqrt {{p_p}} \mathbf{G}_w {\left[ {{{\left( {\bOm + \mathbf{I}_N } \right)}^{ - 1}}} \right]^{1/2}}{\bs{\Phi} ^T} + \mathbf{N},
\end{equation}
which can be further written as
\begin{equation}
\mathbf{Y}_{p,w}= \sqrt {{p_p}} \mathbf{G}_w \mathbf{F}^T + \mathbf{N}.
\end{equation}

The minimum mean-square-error (MMSE) estimate of ${{\bf{\hat G}}_w}$ from $\mathbf{Y}_{p,w}$ is \cite{kay93}
\begin{equation}
{{\bf{\hat G}}_w} = \frac{1}{{\sqrt {{p_p}} }}{\mathbf{Y}_p}{\mathbf{F} ^*}{\bf{\tilde D}},
\end{equation}
where ${\bf{\tilde D}} \triangleq {\left( {\frac{1}{{{p_p}}}{{\bf{D}}^{ - 1}} + {{\bf{I}}_N}} \right)^{ - 1}}$. From \eqref{received pilot}, we can easily get
\begin{equation}\label{gw jian}
{{\bf{\hat G}}_w} =\left( {{{\bf{G}}_w} + \frac{1}{{\sqrt {{p_p}} }}{\bf{W}}} \right){\bf{\tilde D}},
\end{equation}
where $\bf{W} \triangleq \bf{N}\mathbf{F}^*$. Noting that $\mathbf{F}^H\mathbf{F}={\bf I}_N$, the entries of $\bf W$ are i.i.d.~Gaussian random variables with zero-mean and unit-variance. Let $\bs{\Xi} \triangleq \bf{\hat G}-\bf G$ denote the channel estimation error and $\mathbf{\hat A}$ be the linear receiver matrix which depends on $\bhG$. Then, after linear reception, we have
\begin{equation}
{\mathbf{\hat r}} = \sqrt {{p_u}} {\mathbf{\hat A}^H} \left( \mathbf{\hat {G}x}-\bs{\Xi x}\right) + {\mathbf{\hat A}^H}\mathbf{n}.
\end{equation}
As such, the estimated signal for the $n$th user is given by
\begin{equation}\label{imperfect r nth}
{\hat r_n} = \sqrt {{p_u}} \mathbf{\hat a}_n^H\mathbf{\hat Gx} -\sqrt {{p_u}} \mathbf{\hat a}_n^H \bs{\Xi}\mathbf{x}+ \mathbf{\hat a}_n^H\mathbf{n},
\end{equation}
where $\mathbf{\hat a}_n$ is the $n$th column of $\mathbf{\hat A}$, and \eqref{imperfect r nth} can be easily rewritten as
\begin{equation}\label{received signal}
{\hat r_n} \ntsp=\ntsp \sqrt {{p_u}} \mathbf{\hat a}_n^H{\mathbf{\hat g}_n}{x_n} + \sqrt {{p_u}}\ntsp \sum_{i = 1\atop i \ne n}^N{\mathbf{\hat a}_n^H{\mathbf{\hat g}_i}{x_i}} -\sqrt {p_u}\ntsp \sum_{i=1}^N{\mathbf{\hat a}_n^H \bs{\xi_i}\mathbf{x_i}} + \mathbf{\hat a}_n^H\mathbf{n},
\end{equation}
where $\mathbf{\hat g}_i$ and $\bs{\xi}_i$ are the $i$th columns of $\mathbf{\hat G}$ and $\bs{\Xi}$, respectively. Note that the last three terms in \eqref{received signal} correspond to intracell interference, channel estimation error and noise, respectively. According to the classical assumption of worst-case uncorrelated Gaussian noise \cite{Ngo11}, along with the fact that
the variance of elements of estimation error vector $\bs{\xi}_i$ is
\begin{equation}
\bbe\left\{\mathlarger {\left|\right.}{\left[ \bvap  \right]_{mi}}-\bbe\left\{\left[ \bvap  \right]_{mi}\right\}\mathlarger{\left.\right|}^2\right\}={\frac{{{\beta _i}}}{{\left( {1 + {p_p}{\beta _i}} \right)\left( {K_i + 1} \right)}}},
\end{equation}
the achievable uplink rate of the $n$th user can be given by \eqref{imperfect uplink rate} (at the top of the next page).
\begin{figure*}[!t]
\begin{equation}\label{imperfect uplink rate}
{R_{{\sf IP},n}} = \bbe\left\{ {{{\log }_2}\left( {1+ \frac{{{p_u}{{\left| {{\bf{\hat a}}_n^H{{{\bf{\hat g}}}_n}} \right|}^2}}}{{{p_u}\sum_{i = 1\atop i \ne n}^N {{{\left| {{\bf{\hat a}}_n^H{{{\bf{\hat g}}}_i}} \right|}^2}} + \sum_{i = 1}^N {{{\left\|{\bf{\hat a}}_n\right\|^2{\frac{{{p_u\beta _i}}}{{\left( {1 + {p_p}{\beta _i}} \right)\left( {K_i + 1} \right)}}}}} + } {{\left\| {{{{\bf{\hat a}}}_n}} \right\|}^2}}}} \right)} \right\}.
\end{equation}
\hrulefill
\end{figure*}

Similar to the perfect CSI case, the uplink sum rate per cell can be defined as
\begin{equation}\label{imperfect sum rate definition}
C_{\sf IP}=\frac{T-\tau}{T}\sum_{n = 1}^N {{R_{\sf {IP},n}}},
\end{equation}
where $T$ represents the coherence time of the channel, in terms of the number of symbols, during which $\tau$ $(\tau \leq T)$ symbols are used as pilots for channel estimation.

\section{Analysis of Achievable Uplink Rate}\label{sec:achievable uplink rate}
In this section, we derive closed-form expressions for the achievable uplink rates in the large-antenna limit along with tractable approximations that hold for any finite number of antennas. Note that our results are tight and apply for systems with arbitrary-rank Ricean fading channel mean matrices. We also quantify the power-scaling laws in the cases of perfect and imperfect CSI. A key preliminary result is given first.
{ \begin{lemma}\label{lemma 0}
    If \begin{small}$X=\sum_{i=1}^{t_1}{X_i}$\end{small} and \begin{small}$Y=\sum_{j=1}^{t_2}{Y_j}$\end{small} are both sums of nonnegative random variables $X_i$ and $Y_j$, then we get the following approximation
   \begin{small}
   \begin{equation}\label{approximation lemma}
   \bbe \left\{\log_2\left(1+ \frac{X}{Y}\right)\right\} \approx \log_2 \left(1+ \frac{\bbe\left\{X\right\}}{\bbe\left\{Y\right\}}\right),
   \end{equation}
   \end{small}
   \end{lemma}
   \proof
   See Appendix \ref{sec:proof of lemma 0}.
   \endproof

   Note that the approximation in \eqref{approximation lemma} does not require the random variables $X$ and $Y$
to be independent and becomes more accurate as $t_1$ and $t_2$  increase. Thus, in massive MIMO systems, due to the large number of BS antennas, this approximation will be particularly accurate.

\subsection{Perfect CSI}\label{subsec:perfect}
We begin by considering the case with perfect CSI and establishing some key preliminary results which will be useful in deriving our main results.

\begin{lemma}\label{lemma 1}
By the law of large numbers, when $M$ is asymptotically large, we have
\begin{equation}\label{perfect H limit}
\frac{1}{M}\bH^H\bH \xrightarrow{a.s.} \mathbf{I}_N,
\end{equation}
where --a.s. denotes almost sure convergence.
\end{lemma}

\proof See{ Appendix \ref{sec:proof of lemma 1}}.
\endproof

\begin{lemma}\label{lemma 2}
The expectation for the inner product of two same columns in $\bH$ can be found as
\begin{equation}\label{mean 1}
\mathbb{E}{\left\{\left\| {{{\bh}_n}} \right\|^2\right\}} = \mathbb{E}\left\{ {\bf h}_n^H{\bf h}_n\right\} = M ,
\end{equation}
and the expectation of the norm-square of the inner product of any two columns in $\bH$ is given by
{\begin{small}
\begin{numcases}{\bbe{\left\{\left| {{\bh}_n^H{{\bh}_i}} \right|^2\right\}}\ntsp=\ntsp}
{\frac{{2MK_n\ntsp +\ntsp M}}{{{{(K_n \ntsp+\ntsp 1)}^2}}} \ntsp+\ntsp {M^2}}, \ntsp\ntsp\ntsp & \ntsp $i=n$,\label{mean  2}\\
\frac{{{K_n K_i}{\phi ^2_{ni}}\ntsp +\ntsp M\left(K_n\ntsp+\ntsp K_i\right) \ntsp+\ntsp M}}{{{{(K_n \ntsp+\ntsp 1)(K_i \ntsp+\ntsp 1)}}}}, \ntsp\ntsp\ntsp & \ntsp $i \ne n$,\label{mean 3}
\end{numcases}
\end{small}}
\hspace{-4pt}where $\phi_{ni}$ is defined as
\begin{equation}\label{phi definition}
\phi_{ni} \triangleq \frac{{\sin \left( {\frac{M\pi}{2}\left[ {\sin\left({\theta _n}\right) - \sin\left({\theta _i}\right)} \right]} \right)}}{ {{\sin \left( {\frac{\pi}{2}\left[ {\sin\left({\theta _n}\right) - \sin\left({\theta _i}\right)} \right]} \right)}}}.
\end{equation}
\end{lemma}

\proof See{ Appendix \ref{sec:proof of lemma 2}}.
\endproof

\subsubsection{MRC Receivers}
For MRC, the linear receiver matrix is given by
\begin{equation}
\mathbf{A} = \mathbf{G},
\end{equation}
which yields $\mathbf{a}_n = \mathbf{g}_n$. From \eqref{perfect uplink rate}, the achievable rate of the $n$th user is
\begin{small}
\begin{equation}\label{mrc uplink rate}
{R_{{\sf P},n}^{\sf mrc}} = \bbe\left\{ {{{\log }_2}\left( {1 + \frac{{{p_u}{{\left\| {{\mathbf{g}_n}} \right\|}^4}}}{{{p_u}\sum_{i = 1\atop i \ne n}^N {{{\left| {\mathbf{g}_n^H{\mathbf{g}_i}} \right|}^2}}  + {{\left\| {{\mathbf{g}_n}} \right\|}^2}}}} \right)} \right\}.
\end{equation}
\end{small}

Next, we will investigate the power-scaling properties of the uplink rate in \eqref{mrc uplink rate}; this is done by first presenting the exact rate limit in the theorem below.

\begin{theorem}\label{theorem 1}
Using MRC receivers with perfect CSI, if the transmit power of each user is scaled down by a factor of ${1}/{M^\alpha}$, i.e., $p_u={E_u}/{M^\alpha}$ for $\alpha > 0$ and a fixed $E_u$, we have
\begin{equation}\label{mrc limitation}
R_{{\sf P},n}^{\sf mrc} - \dt{R}_{{\sf P},n}^{\sf mrc} \to 0, ~\mbox{as }M \to \infty,
\end{equation}
where $\dt{R}_{{\sf P},n}^{\sf mrc} \triangleq {\log _2}\left( {1 + \frac{{{E_u}{\beta _n}}}{{{M^{\alpha  - 1}}}} }\right)$.
\end{theorem}

\proof
{Let $p_u={E_u}/{M^\alpha}$, where $\alpha >0$. Substituting it into \eqref{mrc uplink rate}, we obtain
\begin{small}
\begin{equation}\label{multiply 1/Ma}
R_{{\sf P},n}^{\sf mrc} = \bbe\left\{ {{{\log }_2}\left( {1 + \frac{{\frac{{{E_u}}}{M^\alpha}{{\left\| {{{\bf{g}}_n}} \right\|}^4}}}{{\frac{{{E_u}}}{M^\alpha}\sum_{i = 1\atop i \ne n}^N {{{\left| {{\bf{g}}_n^H{{\bf{g}}_i}} \right|}^2}}  + {{\left\| {{{\bf{g}}_n}} \right\|}^2}}}} \right)} \right\}.
\end{equation}
\end{small}
\hspace{-5pt}We can also get
\begin{small}
\begin{equation}\label{multiply 1/M^2}
R_{{\sf P},n}^{\sf mrc} = \bbe\left\{ {{{\log }_2}\left( {1 + \frac{{\frac{{{E_u}}}{M^\alpha}\frac{1}{M^2}{{\left\| {{{\bf{g}}_n}} \right\|}^4}}}{{\frac{{{E_u}}}{M^\alpha}\frac{1}{M^2}\sum_{i = 1\atop i \ne n}^N {{{\left| {{\bf{g}}_n^H{{\bf{g}}_i}} \right|}^2}}  + \frac{1}{M^2}{{\left\| {{{\bf{g}}_n}} \right\|}^2}}}} \right)} \right\}.
\end{equation}
\end{small}
\hspace{-5pt}From {\it Lemma \ref{lemma 1}}, it is know that
\begin{small}
\begin{equation}\label{mrc fac1}
\frac{1}{{{M^{\alpha}}}}\frac{1}{M^2}{\left| {{\bf{g}}_n^H{{\bf{g}}_i}} \right|^2}{\rm{ }} = \frac{{\beta _n}{\beta _i}}{M^{\alpha }}{\left| {\frac{1}{M}{\bf{h}}_n^H{{\bf{h}}_i}} \right|^2} \xrightarrow{a.s.} 0, ~\mbox{as }M \to \infty.
\end{equation}
\end{small}
\hspace{-5pt}Then, \eqref{multiply 1/M^2} becomes\footnote{Note that the convergence of logarithms is sure, not almost sure as was shown in \cite{Hoydis13}.}
\begin{equation}\label{mrc simplification convergence}
R_{{\sf P},n}^{\sf mrc} - \bbe\left\{\log_2\left(1+\frac{{{E_u}}}{{{M^\alpha }}}{\left\| {{{\bf{g}}_n}} \right\|^2}\right)\right\}\to 0,~\mbox{as }M \to \infty.
\end{equation}
Using
\begin{equation}
\frac{{{E_u}}}{{{M^\alpha }}}{\left\| {{{\bf{g}}_n}} \right\|^2} = \frac{{{E_u}}}{{{M^{\alpha  - 1}}}}\left| {\frac{1}{M}{\bf{g}}_n^H{{\bf{g}}_n}} \right|,
\end{equation}
and {\it Lemma \ref{lemma 1}} again, it can be easily shown that\footnote{Note that ${{{E_u}{\beta _n}}}/{{{M^{\alpha  - 1}}}}$ is not a constant value and is actually the deterministic equivalent of ${{{E_u}}}{\left\| {{{\bf{g}}_n}} \right\|^2}/{{{M^\alpha }}}$. Therefore, it is a notational abuse to write $\frac{{{E_u}}}{{{M^\alpha }}}{\left\| {{{\bf{g}}_n}} \right\|^2} \xrightarrow{a.s.} \frac{{{E_u}{\beta _n}}}{{{M^{\alpha  - 1}}}}$. }
\begin{equation}\label{mrc inner}
\frac{{{E_u}}}{{{M^\alpha }}}{\left\| {{{\bf{g}}_n}} \right\|^2} - \frac{{{E_u}{\beta _n}}}{{{M^{\alpha  - 1}}}} \xrightarrow{a.s.} 0 , ~\mbox{as }M \to \infty.
\end{equation}
Applying \eqref{mrc inner} in \eqref{mrc simplification convergence} yields the desired result.}
\endproof

{Note that we find a deterministic equivalent for $R_{{\sf P},n}^{\sf mrc}$ in {\it Theorem \ref{theorem 1}} and the value of $\dt{R}_{{\sf P},n}^{\sf mrc}$ varies depending on $\alpha$.} If $\alpha>1$, $\dt{R}_{{\sf P},n}^{\sf mrc}$ will converge to zero, which means that the transmit power of each user has been cut down too much. In contrast, if $\alpha<1$, $\dt{R}_{{\sf P},n}^{\sf mrc}$ will grow without bound, which means that the transmit power of each user can be reduced more to maintain the same performance. {These observations illustrate that only $\alpha =1$ can make $\dt{R}_{{\sf P},n}^{\sf mrc}$ become a fixed non-zero value. Hence, by setting $\alpha=1$ in \eqref{mrc limitation}, we have the following corollary:}
\begin{corollary}\label{corollary 1.1}
With no degradation in the $n$th user's rate, using MRC receivers and perfect CSI, the transmit power of each user can be cut down at most by ${1}/{M}$. Then, the achievable uplink rate becomes
\begin{equation}\label{mrc limitation 1}
R_{{\sf P},n}^{\sf mrc} \to {\log _2}\left( {1 + {{E_u}{\beta _n}} }\right),~\mbox{as }M \to \infty.
\end{equation}
\end{corollary}
\vspace{5pt}

{\it Corollary \ref{corollary 1.1}} shows that when $M$ grows without bound, we can scale down the transmit power proportionally to ${1}/{M}$ to maintain the same rate, which aligns with the conclusion in \cite{Ngo11} for the case of Rayleigh fading. It also indicates that the exact limit of the uplink rate with MRC processing and perfect CSI is irrelevant to the underlying fading model, since the Ricean $K$-factor does not appear in the limiting expression.

The following theorem presents a closed-form approximation for the achievable uplink rate with MRC receivers and perfect CSI. This approximation hold for any finite number of antennas; it can also reveal more effectively the impact of the Ricean $K$-factor on the rate performance.

\begin{theorem}\label{theorem 2}
Using MRC receivers with perfect CSI, the achievable uplink rate of the $n$th user can be approximated as
\begin{small}
\begin{multline}\label{MRC lowBound}
R_{{\sf P},n}^{\sf mrc} \approx \tilde R_{{\sf P},n}^{\sf mrc} = \\{\log _2}\left( {1 + \frac{{{p_u}{\beta _n}\left[ {2MK_n + M + {M^2}{{\left( {K_n + 1} \right)}^2}} \right]}}{{{p_u}(K_n+1)\sum_{i = 1\atop i \ne n}^N {{\beta _i}\Delta_1 }  + M{{\left( {K_n + 1} \right)}^2}}}} \right),
\end{multline}
\end{small}
\hspace{-4pt}where \begin{small}$\Delta_1  \buildrel \Delta \over = \left[{K_n K_i}\phi^2_{ni}+ M(K_n+K_i) + M\right]/(K_i+1)$\end{small}.
\end{theorem}

\proof
Applying {\it Lemma \ref{lemma 0}} on \eqref{mrc uplink rate} yields
\begin{small}
\begin{multline}\label{using inequality}
R_{{\sf P},n}^{\sf mrc} \approx \tilde R_{{\sf P},n}^{\sf mrc}=\\{\log _2}\left( {1 + \frac{{{p_u}\bbe\left\{{{\left\| {{{\bf{g}}_n}} \right\|}^4}\right\}}}{{{p_u}\sum_{i = 1\atop i \ne n}^N {\bbe\left\{{{\left| {{\bf{g}}_n^H{{\bf{g}}_i}} \right|}^2}\right\}}  + \bbe\left\{{{\left\| {{{\bf{g}}_n}} \right\|}^2}\right\}}}} \right).
\end{multline}
\end{small}
\hspace{-4.2pt}The three expectations in \eqref{using inequality} can be evaluated using the results in {\it Lemma \ref{lemma 2}}. Now, with $\bg_n=\beta_n\bh_n$, the substitution of \eqref{mean 1}--\eqref{mean 3} into \eqref{using inequality} leads to the final result.
\endproof

When $K_n=K_i=0$, \eqref{MRC lowBound} reduces to the special case of Rayleigh fading channels:
\begin{equation}\label{perfect rayleigh rate}
\tilde R_{{\sf Rayleigh,P},n}^{\sf mrc}\mathop  = {\log _2}\left( {1 + \frac{{{p_u}{\beta _n}\left( {M + 1} \right)}}{{{p_u}\sum_{i = 1\atop i \ne n}^N {{\beta _i}}  + 1}}} \right).
\end{equation}
It is known that
\begin{equation}\label{perfect MRC rayleigh compare}
\tilde R_{{\sf Rayleigh,P},n}^{\sf mrc}\mathop > {\log _2}\left( {1 + \frac{{{p_u}{\beta _n}\left( {M - 1} \right)}}{{{p_u}\sum_{i = 1\atop i \ne n}^N {{\beta _i}}  + 1}}} \right),
\end{equation}
and the right hand side of \eqref{perfect MRC rayleigh compare} is the uplink rate lower bound with perfect CSI in Rayleigh fading channels given by \cite[Proposition 2]{Ngo11}.

We use the expression in {\it Theorem \ref{theorem 2}} to analyze how the uplink rate changes with the Ricean $K$-factor. The following corollary presents the uplink rate limit when the Ricean $K$-factor grows without bound. To the best of our knowledge, this result is also new.
\begin{corollary}\label{corollary 1}
If for any $n$ and $i$, $K_n=K_i\to\infty$, the approximation in \eqref{MRC lowBound} converges to
\begin{small}
\begin{equation}\label{MRC K limitation}
\tilde R_{{\sf P},n}^{\sf mrc} \to {\log _2}\left({1 + \frac{{{p_u}{\beta _n}{M^2}}}{{{p_u}\sum_{i = 1\atop i \ne n}^N {{\beta _i}{\phi ^2_{ni}}}  + M}}} \right).
\end{equation}
\end{small}
\end{corollary}

It is interesting to note from {\it Corollary \ref{corollary 1}} that as $K$ grows, the uplink rate of MRC will also approach a constant value.
With the power-scaling law uncovered in {\it Corollary \ref{corollary 1.1}}, we now investigate the impact of power reduction on the uplink rate approximation.

\begin{corollary}\label{corollary 2}
With $p_u ={E_u}/{M}$, when $M$ grows without bound while the Ricean $K$-factor is fixed, the result in {\it Theorem \ref{theorem 2}} tends to
\begin{equation}\label{MRC lowBound Limitation}
\tilde R_{{\sf P},n}^{\sf mrc} \to {\log _2}\left( {1 + {E_u}{\beta _n}} \right).
\end{equation}
\end{corollary}

Comparing {\it Corollary \ref{corollary 2}} with {\it Corollary \ref{corollary 1.1}}, we find that when $M$ is large, the uplink rate approximation becomes the same as the exact rate. This is because as $M \to \infty$, things that are random become deterministic and \eqref{approximation lemma} will hold with equality.

\subsubsection{ZF Receivers}
For ZF with $M \geq N+1$, the linear receiver is given by $\mathbf{A}^H\mathbf{G} = {\mathbf{I}_N}$ so that \cite{wang07}
\begin{equation}\label{ZF receiver}
{\bf{a}}_n^H{{\bf{g}}_i} = {\delta _{ni}}.
\end{equation}
Substituting \eqref{ZF receiver} into \eqref{perfect uplink rate} yields \cite{matthaiou11}
\begin{small}
\begin{equation}\label{ZF uplink rate}
R_{{\sf P},n}^{\sf zf} = \mathbb{E}\left\{ {{{\log }_2}\left( {1 + \frac{{{p_u}}}{{{{\left[ {{{\left( {{{\bf{G}}^H}{\bf{G}}} \right)}^{ - 1}}} \right]}_{nn}}}}} \right)} \right\}.
\end{equation}
\end{small}

Similar to the case with MRC receivers, we first quantify the power-scaling law as follows.

\begin{theorem}\label{theorem 3}
Using ZF receivers with perfect CSI, if the transmit power of each user is scaled down by a factor of ${1}/{M^\alpha}$, i.e., $p_u={E_u}/{M^\alpha}$ for $\alpha > 0$ and a fixed $E_u$, we have
\begin{equation}\label{ZF limitation}
R_{{\sf P},n}^{\sf zf} - \dt{R}_{{\sf P},n}^{\sf zf} \to 0, ~\mbox{as }M \to \infty,
\end{equation}
where $\dt{R}_{{\sf P},n}^{\sf zf} \triangleq {\log _2}\left( {1 + \frac{{E_u}{\beta _n}}{M^{\alpha-1}}} \right)$.
\end{theorem}

\proof  With $p_u={E_u}/{M^\alpha}$, we can write \eqref{ZF uplink rate} as
\begin{small}
\begin{equation}
R_{{\sf P},n}^{\sf zf} = \bbe\left\{ {{{\log }_2}\left( {1 + \frac{{{E_u\beta_n}}}{M^\alpha {{{\left[ {{{\left( {{{\bf{H}}^H}{\bf{H}}} \right)}^{ - 1}}} \right]}_{nn}}}}} \right)} \right\},
\end{equation}
\end{small}
\hspace{-5pt}which leads to
\begin{small}
\begin{equation}\label{zf multiply 1/Ma}
R_{{\sf P},n}^{\sf zf}= \bbe\left\{ {{{\log }_2}\left( {1 + \frac{{{E_u\beta_n}}}{M^{\alpha-1} {{{\left[ {{{\left( \frac{1}{M}{{{\bf{H}}^H}{\bf{H}}} \right)}^{ - 1}}} \right]}_{nn}}}}} \right)} \right\}.
\end{equation}
\end{small}
With the aid of {\it Lemma \ref{lemma 1}}, we can easily get the desired result.
\endproof

It is interesting to note from {\it Theorem \ref{theorem 3}} that with increasing $M$, the deterministic equivalent for the uplink rate of ZF receivers, $\dt{R}_{{\sf P},n}^{\sf zf}$, is the same as that of MRC receivers, $\dt{R}_{{\sf P},n}^{\sf mrc}$. Hence, we can draw the conclusion that with no degradation in the $n$th user's rate, in the case of ZF receivers and perfect CSI, the transmit power of each user can be cut down at most by ${1}/{M}$; then, the achievable uplink rate becomes
\begin{equation}
R_{{\sf P},n}^{\sf zf} \to {\log _2}\left( {1 + E_u\beta_n} \right),~\mbox{as }M \to \infty.
\end{equation}


The following theorem gives a closed-form approximation for the uplink rate using ZF receivers with perfect CSI. Note that the steps to derive this expression are more complicated than in the case with MRC receivers, since we need to deal with complex non-central Wishart matrices.

\begin{theorem}\label{theorem 4}
Using ZF receivers with perfect CSI, the achievable uplink rate of the $n$th user can be approximated by
\begin{equation}\label{ZF lowBound}
 R_{{\sf P},n}^{\sf zf} \approx \tilde R_{{\sf P},n}^{\sf zf} = {\log _2}\left( {1 + \frac{{{p_u}{\beta _n}\left( {M - N} \right)}}{{{{\left[ {{\boldsymbol{\hat \Sigma }^{ - 1}}} \right]}_{nn}}}}} \right),
\end{equation}
where
\begin{small}
\begin{align}
\boldsymbol{\hat\Sigma} &\triangleq  \left(\bOm+\mathbf{I}_N\right)^{-1}\\
&+ \frac{1}{M}{\left[ {\bOm {{\left( {\bOm +\mathbf{I}_N } \right)}^{ - 1}}} \right]^{1/2}}{{{\bf{\bar H}}}^H}{\bf{\bar H}}{\left[ {\bOm {{\left( {\bOm +\mathbf{I}_N } \right)}^{ - 1}}} \right]^{1/2}}.
\end{align}
\end{small}
\end{theorem}

\proof By the convexity of $\log_2 \left(1+\frac{1}{x}\right)$ and Jensen's inequality, we obtain the following lower bound on the achievable uplink rate in \eqref{ZF uplink rate}
\begin{equation}
R_{{\sf P},n}^{\sf zf} \geq  {\log _2}\left( {1 + \frac{{{p_u}}}{{\bbe\left\{{{\left[ {{{\left( {{{\bf{G}}^H}{\bf{G}}} \right)}^{ - 1}}} \right]}_{nn}}\right\}}}} \right).\label{ZF using 1/M g}
\end{equation}
With $\bG=\bH\mathbf{D}^{1/2}$, we can rewrite \eqref{ZF using 1/M g} as
\begin{equation}\label{ZF using 1/M}
 R_{{\sf P},n}^{\sf zf} \geq {\log _2}\left( {1 + \frac{{{p_u}{\beta _n}}}{{\bbe\left\{{{\left[ {{{\left( {{{\bf{H}}^H}{\bf{H}}} \right)}^{ - 1}}} \right]}_{nn}}\right\}}}} \right).
\end{equation}

Note that ${{\mathbf{H}}^H}{\bf{H}}$ follows a non-central Wishart distribution, denoted by ${{\bf{H}}^H}{\bf{H}}\sim {\cal W}_N\left( {M,{\mathbf{P}},\bs{\Sigma}}\right)$ \cite{tulino04}, where {\boldmath $\Sigma$} is the covariance matrix of the row vectors of $\mathbf{H}$, i.e., $\boldsymbol{\Sigma} \triangleq \left(\bOm+\mathbf{I}_N\right)^{-1}$ and $\mathbf{P}$ is the mean matrix of $\mathbf{H}$, i.e.,
$$\mathbf{P} \triangleq {\bf{\bar H}}{\left[ {\bOm {{\left( {\bOm +\mathbf{I}_N } \right)}^{ - 1}}} \right]^{1/2}}.$$
 Now, we can approximate ${{\mathbf{H}}^H}{\bf{H}}$ by a central Wishart distribution with covariance matrix as \cite{Steyn72}\footnote{Note that this approximation has been often used in the context of MIMO systems, e.g., \cite{Matthaiou10condition,siriteanu12}.}
\begin{align}
\boldsymbol{\hat\Sigma}  &= \left(\bOm+\mathbf{I}_N\right)^{-1}\\
&+\frac{1}{M}{\left[ {\bOm {{\left( {\bOm +\mathbf{I}_N } \right)}^{ - 1}}} \right]^{1/2}}{{{\bf{\bar H}}}^H}{\bf{\bar H}}{\left[ {\bOm {{\left( {\bOm +\mathbf{I}_N } \right)}^{ - 1}}} \right]^{1/2}}.\label{Sigma jian}
\end{align}

Now, let ${r_n} = \frac{1}{{{{\left[ {{{\left( {{{\bf{H}}^H}{\bf{H}}} \right)}^{ - 1}}} \right]}_{nn}}}}$. Under the above assumption, this is a Chi-squared random variable with distribution \cite{gore02}
\begin{equation}
f({r_n}) = \frac{{{{\left[ {{{\boldsymbol{\hat\Sigma} }^{ - 1}}} \right]}_{nn}}{e^{ - {r_n}{{\left[ {{{\boldsymbol{\hat\Sigma} }^{ - 1}}} \right]}_{nn}}}}}}{{\Gamma \left( {M - N + 1} \right)}}{\left( {{r_n}{{\left[ {{{\boldsymbol{\hat\Sigma} }^{ - 1}}} \right]}_{nn}}} \right)^{M - N}},
\end{equation}
where $\Gamma\left(M\right)=\left(M-1\right)!$ is the gamma function. By expressing the integral $\int_0^\infty  {\frac{1}{{{r_n}}}f\left( {{r_n}} \right)d{r_n}} $, we have\footnote{The first negative moment of $r_n$ exists only if $M>N$ \cite{matthaiou11}.}
\begin{equation}\label{ZF expection}
\bbe\left\{{\left[ {{{\left( {{{\bf{H}}^H}{\bf{H}}} \right)}^{ - 1}}} \right]_{nn}}\right\} = \frac{{{{\left[ {{{\boldsymbol{\hat\Sigma} }^{ - 1}}} \right]}_{nn}}}}{{M - N}}.
\end{equation}
From \eqref{ZF expection} and \eqref{ZF using 1/M}, \eqref{ZF lowBound} can be easily obtained.
\endproof


The following corollary presents the exact limit of the approximation in \eqref{ZF lowBound} for ZF receivers with perfect CSI as the Ricean $K$-factor becomes large.

\begin{corollary}\label{corollary 3}
If for any $n$ and $i$, $K_n=K_i \to \infty$, the approximation in \eqref{ZF lowBound} converges to\footnote{Note that {\it Corollary \ref{corollary 3}} requires $\mathbf{\bar H}$ to be full-rank and this can be guaranteed with great probability in practice, since randomly distributed users are very likely to have different angles of arrival. 
}
\begin{small}
\begin{equation}\label{perfect ZF K}
\tilde R_{{\sf P},n}^{\sf zf} \to {\log _2}\left( {1 + \frac{{{p_u}{\beta _n}\left(M-N\right)}}{{\left[ {\left(\frac{1}{M}{{{\bf{\bar H}}}^H}{\bf{\bar H}}\right)^{-1}} \right]_{nn}}}} \right).
\end{equation}
\end{small}
\end{corollary}

By the discussion after {\it Theorem \ref{theorem 3}}, we know that the transmit power of each user with ZF receivers and perfect CSI can be at most scaled down by ${1}/{M}$. Applying this result into the approximation \eqref{ZF lowBound} gives the following corollary.

\begin{corollary}\label{Corollary 4}
With $p_u ={E_u}/{M}$, when $M$ grows without bound while $K$ is fixed, the result in {\it Theorem \ref{theorem 4}} tends to
\begin{equation}\label{ZF lowBound Limitation}
\tilde R_{{\sf P},n}^{\sf zf} \to {\log _2}\left( {1 + {E_u}{\beta _n}} \right).
\end{equation}
\end{corollary}

\proof
Substituting $p_u={E_u}/{M}$ into \eqref{ZF lowBound}, we obtain
\begin{equation}\label{ZF lowBound multiply 1/M}
\tilde R_{{\sf P},n}^{\sf zf} = {\log _2}\left( {1 + \frac{{{E_u}{\beta _n}\left( {1 - \frac{N}{M}} \right)}}{{{{\left[ {{\bs{\hat \Sigma }^{ - 1}}} \right]}_{nn}}}}} \right).
\end{equation}
The entries of ${\bf{\bar H}}^H{\bf{\bar H}}$ are known as
\begin{equation}\label{Hbar entries}
\left[ {{{{\bf{\bar H}}}^H}{\bf{\bar H}}} \right]_{n,i} =
\begin{cases}
M, & \mbox{if }i=n .\\
\phi_{ni} {e^{j\frac{\left({M - 1}\right)\pi}{2}\left[ {\sin\left({\theta _n}\right) - \sin\left({\theta _i}\right)} \right]}}, & \mbox{if }i \ne n.
\end{cases}
\end{equation}
When $M$ grows large, we asymptotically have
\begin{small}
\begin{multline}
\frac{\phi_{ni}}{M} {e^{j\frac{\left({M - 1}\right)\pi}{2}\left[ {\sin\left({\theta _n}\right) - \sin\left({\theta _i}\right)} \right]}}= \\\frac{{\sin \left( {\frac{M\pi}{2}\left[ {\sin\left({\theta _n}\right) - \sin\left({\theta _i}\right)} \right]} \right)}}{M {{\sin \left( {\frac{\pi}{2}\left[ {\sin\left({\theta _n}\right) - \sin\left({\theta _i}\right)} \right]} \right)}}}{e^{j\frac{\left({M - 1}\right)\pi}{2}\left[ {\sin\left({\theta _n}\right) - \sin\left({\theta _i}\right)} \right]}} \to 0.
\end{multline}
\end{small}
\hspace{-4pt}Therefore, $\frac{1}{M}{\bf{\bar H}}^H{\bf{\bar H}}$ becomes asymptotically an identity matrix. Applying this result into \eqref{Sigma jian} and then combining it with \eqref{ZF lowBound multiply 1/M}, we obtain the desired result.
\endproof

{\it Corollary \ref{Corollary 4}} shows that when $M$ is large, similar to MRC, the approximation for ZF also converges to the exact limit.

\subsection{Imperfect CSI}\label{subsec:imperfect}
In real scenarios, CSI is estimated at the receiver. A typical channel estimation process using pilots is assumed in this paper and has been described in {Section \ref{sec:model}}. Following a similar approach as for the case with perfect CSI, we first give some key preliminary results.

\begin{lemma}\label{lemma 3}
By the law of large numbers, when $M$ is asymptotically large, the inner product of any two columns in the channel matrix $\bhG$ can be found as
\begin{numcases}{\frac{1}{M}{\bf{\hat g}}_n^H{{{\bf{\hat g}}}_i} \xrightarrow{a.s.}}
\frac{{{\beta _n}}}{{K_n + 1}}\left( {K_n + \eta_n} \right), & \mbox{if }$i=n$,\label{imperfect largeNumberLaw 1}\\
0, & \mbox{if }$i \ne n$,\label{imperfect largeNumberLaw 2}
\end{numcases}
where $\eta_n$ denotes the $n$th column of $\mathbf{\tilde D}$, i.e.,
\begin{equation}\label{eta definition}
\eta_n \triangleq \frac{{{p_p}{\beta_n}}}{{1 + {p_p}{\beta_n}}}.
\end{equation}
\end{lemma}
\proof See {Appendix \ref{sec:proof of lemma 3}}.
\endproof


\begin{lemma}\label{lemma 4}
The expectation for the inner product of two same columns in $\bhG$ can be found as
\begin{equation}
\bbe\left\{\left\| {\bhg_n} \right\|^2\right\}= \bbe\left\{{{\bf{\hat g}}_n^H{{{\bf{\hat g}}}_n}}\right\} = {\beta _n}\left( {\frac{{MK_n}}{{K_n + 1}} + \frac{{M{\eta _n}}}{{K_n + 1}}} \right),\label{MRC expection 1}
\end{equation}
 and the expectation for the norm-square of the inner product of any two columns in $\bhG$ is expressed as \eqref{MRC expection 3} (at the top of the next page).
\begin{figure*}[!t]
\begin{small}
\begin{equation}\label{MRC expection 3}
{\bbe \left\{{\left| {{\bf{\hat g}}_n^H{{{\bf{\hat g}}}_i}} \right|^2}\right\}\ntsp=\ntsp}
\begin{cases}
\dfrac{{\beta _n^2}}{{{{\left( {K_n + 1} \right)}^2}}}\left[ {{M^2}{K_n^2}\ntsp + \ntsp\left( {2MK_n + 2M^2{K_n}} \right){\eta _n}\ntsp + \ntsp\left( {{M^2} + M} \right)\eta _n^2} \right], & \text{if $i = n$}\\[5pt]
\dfrac{{{\beta _n}{\beta _i}}}{{{{\left( {K_n + 1} \right)\left(K_i+1\right)}}}}\left[ {{K_nK_i}{\phi ^2_{ni}} \ntsp+\ntsp MK_i{\eta _n} + MK_n{\eta _i} + M{\eta _n}{\eta _i}} \right] , &\text{if $i \ne n$}
\end{cases}
\end{equation}
\end{small}
\hrulefill
\end{figure*}
\end{lemma}

\proof See {Appendix \ref{sec:proof of lemma 4}}.
\endproof

\subsubsection{MRC Receivers}
For MRC, we have $\mathbf{\hat a}_n=\mathbf{\hat g}_n$. As a result, from \eqref{imperfect uplink rate}, the uplink rate of the $n$th user with MRC receivers can be written as
\begin{small}
\begin{multline}\label{MRC uplink rate}
R_{{\sf IP},n}^{\sf mrc} =\\ \bbe \ntsp \left\{\ntsp {{{\log }_2}\ntsp \left( \ntsp{1\ntsp +\ntsp \frac{{{p_u}{{\left\| {{{{\bf{\hat g}}}_n}} \right\|}^4}}}{{{p_u}\sum_{i = 1\atop i \ne n}^N \ntsp{{{\left| {{\bf{\hat g}}_n^H \nmsp {{{\bf{\hat g}}}_i}} \right|}^2}}  \ntsp+\ntsp \left[ \nmsp \sum_{i = 1}^N\ntsp{\frac{{{p_u\beta _i}}}{{\left( {1 + {p_p}{\beta _i}} \right)\left( {K_i+ 1} \right)}}}\ntsp+\ntsp 1 \nmsp \right] \ntsp{{\left\| {{{{\bf{\hat g}}}_n}} \right\|}^2}}}} \ntsp\right)} \ntsp\right\}.
\end{multline}
\end{small}

Same as before, we will investigate the power-scaling law but this time with imperfect CSI. We will first derive the exact uplink rate and then an approximation for further analysis.

\begin{theorem}\label{theorem 5}
Using MRC receivers with imperfect CSI from MMSE estimation, if the transmit power of each user is scaled down by a factor of ${1}/{M^\alpha}$, i.e., $p_u={E_u}/{M^\alpha}$ for $\alpha > 0$ and a fixed $E_u$, we have
\begin{equation}\label{mrc imperfect limit}
R_{{\sf IP},n}^{\sf mrc} - \dt{R}_{{\sf IP},n}^{\sf mrc} \to 0,~\mbox{as }M\to\infty,
\end{equation}
where $\dt{R}_{{\sf IP},n}^{\sf mrc} \triangleq {\log _2}\left( {1 + \left[ {\frac{{{E_u}{\beta _n}K_n}}{{{M^{\alpha  - 1}}\left( {K_n + 1} \right)}} + \frac{{\tau E_u^2\beta _n^2}}{{{M^{2\alpha  - 1}}\left( {K_n + 1} \right)}}} \right]} \right)$.
\end{theorem}
\proof
{Let ${p_u} = {E_u}/{M^\alpha}$, where $\alpha > 0$. Substituting it into \eqref{MRC uplink rate} yields \eqref{MRC scaling-down pu} (at the top of the next page).
\begin{figure*}[!t]
\begin{small}
\begin{equation}\label{MRC scaling-down pu}
R_{{\sf IP},n}^{\sf mrc} = \bbe \left\{ {{{\log }_2} \left( {1 + \frac{{{\frac{E_u}{M^\alpha}}{{\left\| {{{{\bf{\hat g}}}_n}} \right\|}^4}}}{{{\frac{E_u}{M^\alpha}}\sum_{i = 1\atop i \ne n}^N {{{\left| {{\bf{\hat g}}_n^H{{{\bf{\hat g}}}_i}} \right|}^2}}  + \left[\sum_{i = 1}^N{\frac{{{\frac{E_u}{M^\alpha}\beta _i}}}{{\left( {1 + {\tau \frac{E_u}{M^\alpha}}{\beta _i}} \right)\left( {K_i+ 1} \right)}}}+1\right] {{\left\| {{{{\bf{\hat g}}}_n}} \right\|}^2}}}} \right)} \right\}.
\end{equation}
\end{small}
\hrulefill
\end{figure*}
Let the numerator and denominator of \eqref{MRC scaling-down pu} be multiplied by $1/M^2$, we obtain \eqref{MRC scaling-down pu 1/M^2} (at the top of the next page).
\begin{figure*}[!t]
\begin{small}
\begin{equation}\label{MRC scaling-down pu 1/M^2}
R_{{\sf IP},n}^{\sf mrc} = \bbe \left\{ {{{\log }_2} \left( {1 + \frac{{{\frac{E_u}{M^\alpha}\frac{1}{M^2}}{{\left\| {{{{\bf{\hat g}}}_n}} \right\|}^4}}}{{{\frac{E_u}{M^\alpha}\frac{1}{M^2}}\sum_{i = 1\atop i \ne n}^N {{{\left| {{\bf{\hat g}}_n^H{{{\bf{\hat g}}}_i}} \right|}^2}}  + \frac{1}{M^2}\left[\sum_{i = 1}^N{\frac{{{\frac{E_u}{M^\alpha}\beta _i}}}{{\left( {1 + {\tau \frac{E_u}{M^\alpha}}{\beta _i}} \right)\left( {K_i+ 1} \right)}}}+1\right] {{\left\| {{{{\bf{\hat g}}}_n}} \right\|}^2}}}} \right)} \right\}.
\end{equation}
\end{small}
\hrulefill
\end{figure*}
When $M \to \infty$, using \eqref{imperfect largeNumberLaw 2} in {\it Lemma \ref{lemma 3}}, it can be easily obtained that
\begin{equation}\label{imperfect fac1}
\frac{{{E_u}}}{{{M^{\alpha} }}}\frac{1}{M^2}{\left| {{\bf{\hat g}}_n^H{{{\bf{\hat g}}}_i}} \right|^2} = \frac{{{E_u}}}{{{M^{\alpha}}}}{\left| {\frac{1}{M}{\bf{\hat g}}_n^H{{{\bf{\hat g}}}_i}} \right|^2} \xrightarrow{a.s.} 0,
\end{equation}
and with
\begin{equation}
  \sum_{i = 1}^N{\frac{{{\frac{E_u}{M^\alpha}\beta _i}}}{{\left( {1 + {\tau \frac{E_u}{M^\alpha}}{\beta _i}} \right)\left( {K_i+ 1} \right)}}}+1 \to 1,~\mbox{as }M \to \infty,
\end{equation}
we can further simplify \eqref{MRC scaling-down pu 1/M^2} as
\begin{equation}
R_{{\sf IP},n}^{\sf mrc} - \mathbb E\left\{ {{{\log }_2}\left( {1 + \frac{{{E_u}}}{{{M^\alpha }}}{{\left\| {{{{\bf{\hat g}}}_n}} \right\|}^2}} \right)} \right\} \to 0,~\mbox{as }M \to \infty.
\end{equation}
The remaining task is to evaluate the limit of $\frac{1}{M^\alpha}{{\left\| {{{{\bf{\hat g}}}_n}} \right\|}^2}$. Using \eqref{imperfect largeNumberLaw 1} in {\it Lemma \ref{lemma 3}}, along with the fact that
\begin{equation}
\frac{{{E_u}}}{{{M^\alpha }}}{\left\| {{\bhg_n}} \right\|^2} = \frac{{{E_u}}}{{{M^{\alpha  - 1}}}}\left| {\frac{1}{M}{\bhg}_n^H{{\bhg}_n}} \right|,
\end{equation}
and $p_p=\tau p_u=\tau E_u/M^\alpha$, we get
\begin{small}
\begin{multline}\label{mrc limit simplify 1}
\frac{{{E_u}}}{{{M^\alpha }}}{\left\| {{{{\bf{\hat g}}}_n}} \right\|^2}  - \\\left[ \frac{{{E_u}{\beta _n}K_n}}{{{M^{\alpha  - 1}}\left( {K_n + 1} \right)}} + \frac{{\tau E_u^2\beta _n^2}}{{{M^{2\alpha  - 1}}\left( {K_n + 1} \right)\left( {1 + \frac{\tau {E_u}{\beta _n}}{M^\alpha }} \right)}}\right] \xrightarrow{a.s.} 0.
\end{multline}
\end{small}
\hspace{-5pt}Since $\alpha>0$, ${1}/{M^\alpha} \to 0$ and we can further simplify \eqref{mrc limit simplify 1} to complete the proof.}
\endproof

It is important to note from {\it Theorem \ref{theorem 5}} that the value of the deterministic equivalent for $R_{{\sf IP},n}^{\sf mrc}$ is dependent on both the Ricean $K$-factor and the scaling parameter $\alpha$, which can be more precisely described as follows. When $K_n=0$, with increasing $M$, only $\alpha=0.5$ can make $\dt{R}_{{\sf IP},n}^{\sf mrc}$ approach a non-zero constant value. Otherwise, it will tend to zero with $\alpha>0.5$ and grow without bound with $\alpha<0.5$, which will change the rate performance. For the case $K_n\ne 0$, we find that only $\alpha=1$ can lead $\dt{R}_{{\sf IP},n}^{\sf mrc}$ to a fixed value, and when $\alpha>1$ and $\alpha<1$, it will approximately become zero and $\infty$, respectively. {These observations can be summarized as the power-scaling law
in the following corollary.}

\begin{corollary}\label{corollary 2.1}
With no degradation in the $n$th user's rate, using MRC receivers with imperfect CSI and for a fixed $E_u$, we obtain the following power-scaling law. When the $n$th user's Ricean $K$-factor is zero, we can at most scale down the transmit power to ${p_u} = {E_u}/{\sqrt M}$ and the achievable uplink rates becomes
\begin{equation}\label{mrc imperfect k0 limit}
R_{{\sf IP},n}^{\sf mrc} \to {\log _2}\left( {1 + \tau E_u^2\beta _n^2} \right),~\mbox{as }M \to \infty.
\end{equation}
On the other hand, with a non-zero Ricean $K$-factor, the transmit power can be at most scaled down to ${p_u} = {E_u}/{M}$ and we have
\begin{equation}\label{mrc imperfect k-ne-0 limit}
R_{{\sf IP},n}^{\sf mrc} \to {\log _2}\left( {1 + \frac{{K_n{E_u}{\beta _n}}}{{K_n + 1}}} \right), ~\mbox{as }M \to \infty.
\end{equation}
\end{corollary}

{\it Corollary \ref{corollary 2.1}} reveals that with imperfect CSI, the amount of power reduction achievable on the $n$th user depends on $K_n$. For Rayleigh fading channels, i.e., $K_n=0$, the transmit power can be at most cut down by a factor of ${1}/{\sqrt M}$, which agrees with the result in \cite[Proposition 5]{Ngo11}; however, for channels with LOS components, i.e., $K_n \ne 0$, the transmit power can be cut down by ${1}/{M}$. This is reasonable since LOS propagation reduces fading fluctuations, thereby increasing the received signal-to-interference-plus-noise ratio. As a consequence, with no reduction in the rate performance, the transmit power for non-zero Ricean $K$-factor can be cut down more aggressively. 
 The following theorem presents a closed-form approximation for the uplink rate using MRC with imperfect CSI.

\begin{theorem}\label{theorem 6}
Using MRC receivers with imperfect CSI from MMSE estimation, the achievable uplink rate of the $n$th user can be approximated by \eqref{MRC approximation} (at the top of the next page),
\begin{figure*}[!t]
\begin{small}
\begin{equation}\label{MRC approximation}
 R_{{\sf IP},n}^{\sf mrc} \approx \tilde R_{{\sf IP},n}^{\sf mrc} =\\
 {\log _2}\left( {1 + \frac{{{p_u}{\beta _n}\left[ {{M^2}{K_n^2} + \left( {2MK_n + 2{M^2}K_n} \right){\eta _n} + \left( {M + {M^2}} \right)\eta _n^2} \right]}}{{{p_u}\left(K_n+1\right)\sum_{i = 1\atop i \ne n}^N {{\beta _i}\Delta_2}  + M{p_u}{\beta _n}\frac{{K_n + {\eta _n}}}{{1 + {\beta _n}{p_p}}} + M\left( {K_n + 1} \right)\left( {K_n + {\eta _n}} \right)}}} \right).
\end{equation}
\end{small}
\hrulefill
\end{figure*}
where \begin{small}$\Delta_2 \triangleq \left[ {{K_nK_i}{\phi ^2_{ni}} + M{\eta _n}\left( {K_i + 1} \right) + MK_n} \right]/(K_i+1)$\end{small}.
\end{theorem}

\proof
Follow the same procedure as {\it Theorem 2}. After utilizing {\it Lemma \ref{lemma 0}} and \eqref{MRC expection 1}--\eqref{MRC expection 3}, we can obtain the desired result after some simplifications.
\endproof

Note that
when $K_n=K_i=0$, $\tilde R_{{\sf IP},n}^{\sf mrc}$ reduces to the special case of Rayleigh fading channel. After performing some simplifications, we have
\begin{small}
\begin{multline}\label{imperfect rayleigh rate}
\tilde R_{{\sf Rayleigh,IP},n}^{\sf mrc}= \\{\log _2}\ntsp \left( \ntsp {1 \ntsp +\ntsp \frac{{\tau p_u^2\beta _n^2\left( {M + 1} \right)}}{{{p_u}\left( {\tau {p_u}{\beta _n} + 1} \right)\sum_{i = 1\atop i \ne n}^N  {{\beta _i}}  + \left( {\tau  + 1} \right){p_u}{\beta _n} + 1}}} \right).
\end{multline}
\end{small}
It is known that
\begin{small}
\begin{multline}\label{imperfect MRC rayleigh compare}
\tilde R_{{\sf Rayleigh,IP},n}^{\sf mrc} > \\{\log _2}\ntsp \left( \ntsp {1\ntsp + \ntsp\frac{{\tau p_u^2\beta _n^2\left( {M - 1} \right)}}{{{p_u}\left( {\tau {p_u}{\beta _n} + 1} \right)\sum_{i = 1\atop i \ne n}^N  {{\beta _i}}  + \left( {\tau  + 1} \right){p_u}{\beta _n} + 1}}} \right),
\end{multline}
\end{small}
\hspace{-4pt}and the right hand side of \eqref{imperfect MRC rayleigh compare} is the uplink rate lower bound with imperfect CSI in Rayleigh fading channels given by \cite[Proposition 6]{Ngo11}.

Now, we consider the particular case when the Ricean $K$-factor grows infinite for the approximation in {\it Theorem \ref{theorem 6}}.

\begin{corollary}
If for any $n$ and $i$, $K_n=K_i \to \infty$, the approximation in  \eqref{MRC approximation} converges to
\begin{small}
\begin{equation}
\tilde R_{{\sf IP},n}^{\sf mrc} \to  {\log _2} \left( {1+ \frac{{{p_u}{\beta _n}{M^2}}}{{{p_u}\sum_{i = 1\atop i \ne n}^N {{\beta _i}{\phi ^2_{ni}}}  + M}}} \right).
\end{equation}
\end{small}
\end{corollary}

This conclusion indicates that when the Ricean $K$-factor grows large, the approximation of the uplink rate using MRC with imperfect CSI will approach a fixed value, which is the same as the constant value in the case of perfect CSI given by \eqref{MRC K limitation}. That is, in a purely deterministic channel, the approximation for MRC receivers will tend to the same constant value regardless of the CSI quality at the BS.

With the power-scaling law already derived in {\it Corollary \ref{corollary 2.1}}, we now perform the same analysis but on the uplink rate approximation in {\it Theorem \ref{theorem 6}}.

\begin{corollary}
If $p_u={E_u}/{\sqrt M}$ for a fixed $E_u$, when $M$ grows without bound, the limit of \eqref{MRC approximation} will exist only when $K_n=0$ and \eqref{MRC approximation} becomes
\begin{equation}
\tilde R_{{\sf IP},n}^{\sf mrc} \to {\log _2}\left( {1 + \tau E_u^2\beta _n^2} \right),~\mbox{as }M\to\infty.
\end{equation}
 If $p_u={E_u}/{M}$ for a fixed $E_u$, when $M$ grows without bound, \eqref{MRC approximation} will converge to a non-zero constant value only when $K_n \ne 0$ and \eqref{MRC approximation} becomes
\begin{equation}
\tilde R_{{\sf IP},n}^{\sf mrc} \to {\log _2}\left( {1 + \frac{{K_n{E_u}{\beta _n}}}{{K_n + 1}}} \right),~\mbox{as }M\to\infty.
\end{equation}
\end{corollary}

\proof
With the fact that
\begin{equation}
\frac{1}{\sqrt M}\phi^2_{ni} \to 0,~\mbox{as }M\to\infty,
\end{equation}
the result then follows by performing some basic algebraic manipulations.
\endproof

Again, it is noted that when $M \to \infty$, the approximation of the uplink rate becomes identical to the exact limit obtained from {\it Corollary \ref{corollary 2.1}}.

\subsubsection{ZF Receivers}
For ZF receivers, we have ${\bf{\hat a}}_n^H{{\bf{\hat g}}_i} = {\delta _{ni}}$. According to \eqref{imperfect uplink rate}, the achievable uplink rate of ZF receivers with imperfect CSI is given by
\begin{small}
\begin{multline}\label{ZF imperfect uplink rate original}
{R_{{\sf IP},n}^{\sf zf}} \ntsp =\\ \bbe \left\{{{{\log }_2}\ntsp \left(\ntsp {1 + \frac{{{p_u}}}{{\left( {\sum\nolimits_{i = 1}^N {\frac{{{p_u}{\beta _i}}}{{\left( {1 + {p_p}{\beta _i}} \right)\left( {K_i + 1} \right)}}}  + 1} \right)\left[ {{{{\bf{\hat G}}}^H}{\bf{\hat G}}} \right]_{nn}^{ - 1}}}} \right)}\right\}.
\end{multline}
\end{small}

Similar to the case with MRC receivers, we next investigate the power-scaling law for ZF receivers.

\begin{theorem}\label{theorem 7}
Using ZF receivers with imperfect CSI from MMSE estimation, if the transmit power of each user is scaled down by a factor of ${1}/{M^\alpha}$, i.e., $p_u={E_u}/{M^\alpha}$ for $\alpha > 0$ and a fixed $E_u$, we have
\begin{equation}
R_{{\sf IP},n}^{\sf zf} - \dt{R}_{{\sf IP},n}^{\sf zf} \to 0 ,~\mbox{as }M \to \infty,
\end{equation}
where $\dt{R}_{{\sf IP},n}^{\sf zf} \triangleq \log_2\left(1+\left[\frac{{{E_u}{\beta _n}K_n}}{{{M^{\alpha  - 1}}\left( {K_n + 1} \right)}} + \frac{{\tau E_u^2\beta _n^2}}{{{M^{2\alpha  - 1}}\left( {K_n + 1} \right)}}\right]\right)$.
\end{theorem}

\proof
Let $p_u={E_u}/{M^\alpha}$, where $\alpha >0$. Then, \eqref{ZF imperfect uplink rate original} becomes
\begin{small}
\begin{multline}\label{zf imperfect 1/Ma}
{R_{{\sf IP},n}^{\sf zf}} =\\ \bbe\left\{\ntsp {{{\log }_2}\ntsp \left(\ntsp {1\ntsp +\ntsp \frac{{{E_u}}}{{\left( {\sum\nolimits_{i = 1}^N {\frac{{{\frac{E_u}{M^\alpha}}{\beta _i}}}{{\left( {1 + {\tau\frac{E_u}{M^\alpha}}{\beta _i}} \right)\left( {K_i + 1} \right)}}} \ntsp +\ntsp 1} \right){M^{\alpha \nmsp - \nmsp 1}}\left[ {\frac{1}{M}{{{\bhG}}^H}{\bhG}} \right]_{nn}^{ - 1}}}} \right)}\ntsp \right\}.
\end{multline}
\end{small}
\hspace{-4pt}By {\it Lemma {\ref {lemma 3}}}, we know that
\begin{equation}
\frac{1}{M}{{{\bf{\hat G}}}^H}{\bf{\hat G}} \xrightarrow{a.s.} \bs{\Psi} ,
\end{equation}
where $\bs{\Psi}$ is a diagonal matrix containing $\left[ {\frac{{{\beta _1}}}{{K_1 + 1}}\left( {K_1 \ntsp+\ntsp {\eta _1}} \right),\cdots ,\frac{{{\beta _n}}}{{K_n + 1}}\left( {K_n\ntsp +\ntsp {\eta _n}} \right), \cdots ,\frac{{{\beta _N}}}{{K_N + 1}}\left( {K_N \ntsp+\ntsp {\eta _N}} \right)} \right]$ along its main diagonal. The inverse matrix of $\bs{\Psi}$ can be easily obtained by calculating the reciprocals of its main diagonal. As a consequence, using the fact $\eta_n=p_p \beta_n/\left(1+p_p \beta_n\right)$, we have
\begin{small}
\begin{multline}\label{zf imperfect denominator }
{{M^{\alpha  - 1}}\left[ {\frac{1}{M}{{{\bf{\hat G}}}^H}{\bf{\hat G}}} \right]_{nn}^{ - 1}}  -\\ \left[\frac{K_n\beta_n}{{{M^{\alpha  - 1}}\left( {K_n + 1} \right)}} + \frac{{\tau p_p \beta _n^2}}{{{M^{\alpha  - 1}}\left( {K_n + 1} \right)\left(1+p_p \beta_n \right)}}\right]^{-1} \xrightarrow{a.s.} 0.
\end{multline}
\end{small}
\hspace{-5pt}Due to $\alpha>0$, as $M \to \infty$, the substitution of $p_p=\tau p_u=\tau E_u/M^\alpha$ into \eqref{zf imperfect denominator } yields
\begin{small}
\begin{multline}\label{zf imperfect denominator simplify}
{{M^{\alpha  - 1}}\left[ {\frac{1}{M}{{{\bf{\hat G}}}^H}{\bf{\hat G}}} \right]_{nn}^{ - 1}} -\\ \left[\frac{K_n\beta_n}{{{M^{\alpha  - 1}}\left( {K_n + 1} \right)}} + \frac{{\tau E_u\beta _n^2}}{{{M^{2\alpha  - 1}}\left( {K_n + 1} \right)}}\right]^{-1} \xrightarrow{a.s.} 0,
\end{multline}
\end{small}
\hspace{-5pt}and we can also easily get
\begin{small}
\begin{equation}\label{zf cof tend}
{\sum\nolimits_{i = 1}^N {\frac{{{\frac{E_u}{M^\alpha}}{\beta _i}}}{{\left( {1 + {\tau\frac{E_u}{M^\alpha}}{\beta _i}} \right)\left( {K_i + 1} \right)}}}  + 1} \to 1,~\mbox{as }M \to \infty.
\end{equation}
\end{small}
\hspace{-5pt}The desired result is obtained by substituting \eqref{zf imperfect denominator simplify} and \eqref{zf cof tend} into \eqref{zf imperfect 1/Ma}.
\endproof

From {\it Theorem \ref{theorem 7}}, it is worth pointing out that with imperfect CSI and increasing $M$, the deterministic equivalent for the uplink rate of ZF receivers becomes the same as that of MRC receivers in \eqref{mrc imperfect limit}, which agrees with the conclusion in the case of perfect CSI. Therefore, we can easily obtain the conclusion that with no degradation in the $n$th user's rate performance, using ZF receivers with imperfect
CSI for a fixed $E_u$, when $K_n=0$,
we can at most scale down the transmit power of each user to ${1}/{\sqrt M}$ and the achievable uplink rates becomes
\begin{equation}\label{zf imperfect k0 limit}
R_{{\sf IP},n}^{\sf zf} \to {\log _2}\left( {1 + \tau E_u^2\beta _n^2} \right),~\mbox{as }M \to \infty.
\end{equation}
On the other hand, with a non-zero $K_n$, the transmit power of each user can be at most scaled down to $p_u=E_u/M$ and we have
\begin{equation}\label{zf imperfect k-ne-0 limit}
R_{{\sf IP},n}^{\sf zf} \to {\log _2}\left( {1 + \frac{{K_n{E_u}{\beta _n}}}{{K_n + 1}}} \right),~\mbox{as }M \to \infty.
\end{equation}

It is also important to point out that for both perfect and imperfect CSI, the exact limit of the uplink rate of ZF receivers equals that of MRC receivers, which is consistent with the conclusion given in \cite{Ngo11}. The following theorem presents a closed-form approximation for the achievable uplink rate using ZF receivers with imperfect CSI.

\begin{theorem}\label{Theorem 8}
Using ZF receivers with imperfect CSI from MMSE estimation, the achievable uplink rate of the $n$th user can be approximated by
\begin{small}
\begin{equation}\label{imperfect ZF lowBound}
R_{{\sf IP},n}^{\sf zf} \ntsp \approx \ntsp \tilde R_{{\sf IP},n}^{\sf zf}\ntsp =\ntsp {\log _2}\ntsp\left(\ntsp {1 \ntsp +  \ntsp \frac{{{p_u}{\beta _n}\left( {M - N} \right)}}{\left(\ntsp{\sum\nolimits_{i = 1}^N \ntsp{\frac{{{p_u}{\beta _i}}}{{\left( {1 + {p_p}{\beta _i}} \right)\left( {K_i + 1} \right)}}} \ntsp +\ntsp 1\ntsp}\right){{{\left[ {{\boldsymbol{\tilde \Sigma }^{ - 1}}} \right]}_{nn}}}}} \ntsp\right),
\end{equation}
\end{small}
\hspace{-4pt}where
\begin{small}
\begin{equation}\label{imperfect sigma}
\boldsymbol{\tilde \Sigma}  \triangleq {\bs{\Lambda}} + \frac{1}{M}{\left[ {\bOm {{\left( {\bOm +\mathbf{I}_N } \right)}^{ - 1}}} \right]^{1/2}}{{{\bf{\bar H}}}^H}{\bf{\bar H}}{\left[ {\bOm {{\left( {\bOm +\mathbf{I}_N } \right)}^{ - 1}}} \right]^{1/2}},
\end{equation}
\end{small}
\hspace{-15pt}and $\bs{\Lambda}$ is a diagonal matrix consisting of $\left[\frac{\eta_1}{K_1+1},\cdots,\frac{\eta_n}{K_n+1},\cdots,\frac{\eta_N}{K_N+1}\right]$.
\end{theorem}

\proof
Applying Jensen's inequality in \eqref{ZF imperfect uplink rate original}, we get
\begin{small}
\begin{multline}\label{zf using approximation}
R_{{\sf IP},n}^{\sf zf} \ge \\ {{{\log }_2}\ntsp \left(\ntsp {1 + \frac{{{p_u}}}{{\left( {\sum\nolimits_{i = 1}^N {\frac{{{p_u}{\beta _i}}}{{\left( {1 + {p_p}{\beta _i}} \right)\left( {K_i + 1} \right)}}}  + 1} \right)\bbe\left[ {{{{\bf{\hat G}}}^H}{\bf{\hat G}}} \right]_{nn}^{ - 1}}}} \right)}.
\end{multline}
\end{small}

The proof then follows the same procedure as in {\it Theorem \ref{theorem 4}}.
\endproof


\begin{corollary}\label{Corollary 7}
If for any $n$ and $i$, $K_n=K_i \to \infty$, the approximation in  \eqref{imperfect ZF lowBound} converges to
\begin{small}
\begin{equation}
\tilde R_{{\sf IP},n}^{\sf zf} \to {\log _2}\left( {1 + \frac{{{p_u}{\beta _n}\left(M-N\right)}}{{\left[ {\frac{1}{M}{{{\bf{\bar H}}}^H}{\bf{\bar H}}} \right]_{nn}^{ - 1}}}} \right).
\end{equation}
\end{small}
\end{corollary}

Note that the fixed value in this case is the same as the limit in {\it Corollary \ref{corollary 3}} given by \eqref{perfect ZF K}, which shows that for both MRC and ZF receivers, as the Ricean $K$-factor grows without bound, the approximations of the uplink rate will tend to the same fixed value regardless of the CSI quality.

\begin{corollary}\label{corollary 8}
If $p_u={E_u}/{\sqrt M}$ for a fixed $E_u$, when $M$ grows without bound, the limit of \eqref{imperfect ZF lowBound} will exist only when $K_n=0$ and \eqref{imperfect ZF lowBound} becomes
\begin{equation}
\tilde R_{{\sf IP},n}^{\sf zf} \to {\log _2}\left( {1 + \tau E_u^2\beta _n^2} \right),~\mbox{as }M\to\infty.
\end{equation}
If $p_u={E_u}/{M}$ for a fixed $E_u$, when $M$ grows without bound, \eqref{imperfect ZF lowBound} will converge to a non-zero constant value only when $K_n \ne 0$ and \eqref{imperfect ZF lowBound} becomes
\begin{equation}
\tilde R_{{\sf IP},n}^{\sf zf} \to {\log _2}\left( {1 + \frac{{K_n{E_u}{\beta _n}}}{{K_n + 1}}} \right),~\mbox{as }M\to\infty.
\end{equation}
\end{corollary}
\proof
From the proof of {\it Corollary \ref{Corollary 4}}, we know that when $M$ is large, $\frac{1}{M}{\bf{\bar H}}^H{\bf{\bar H}}$ becomes asymptotically an identity matrix. With the fact that
\begin{small}
\begin{equation}\label{cof to 1}
\sum\nolimits_{i = 1}^N {\frac{{{p_u}{\beta _i}}}{{\left( {1 + {p_p}{\beta _i}} \right)\left( {K_i + 1} \right)}}}  + 1 \to 1, ~\mbox{as } M \to \infty,
\end{equation}
\end{small}
\hspace{-4pt}the result then follows by some basic algebraic manipulations.
\endproof

It is interesting to note from {\it Corollary \ref{corollary 8}} that for both perfect and imperfect CSI, as $M$ grows large, the exact limit equals the approximating expression regardless of the type of the receiver.

\section{Numerical Results}\label{sec:numerical results}

\begin{figure}
\centering
\includegraphics[scale=0.35]{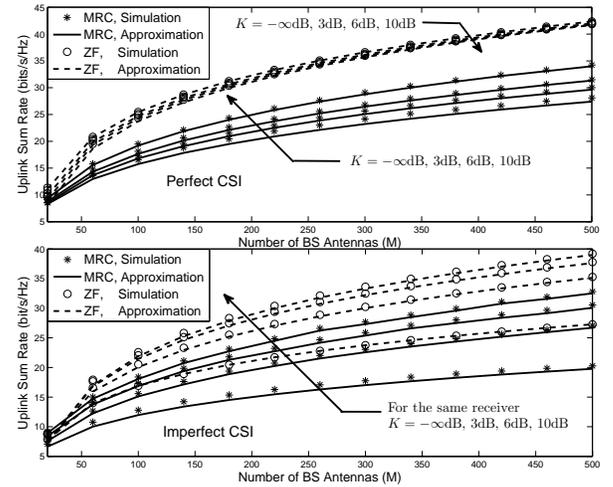}
\caption{Uplink sum rate per cell versus the number of BS antennas, with $N=10$ users and transmit power $p_u=10$dB.}\label{fig:fig 1}
\setlength{\abovecaptionskip}{0pt}
\setlength{\belowcaptionskip}{0pt}
\end{figure}

\begin{figure}
\centering
\includegraphics[scale=0.37]{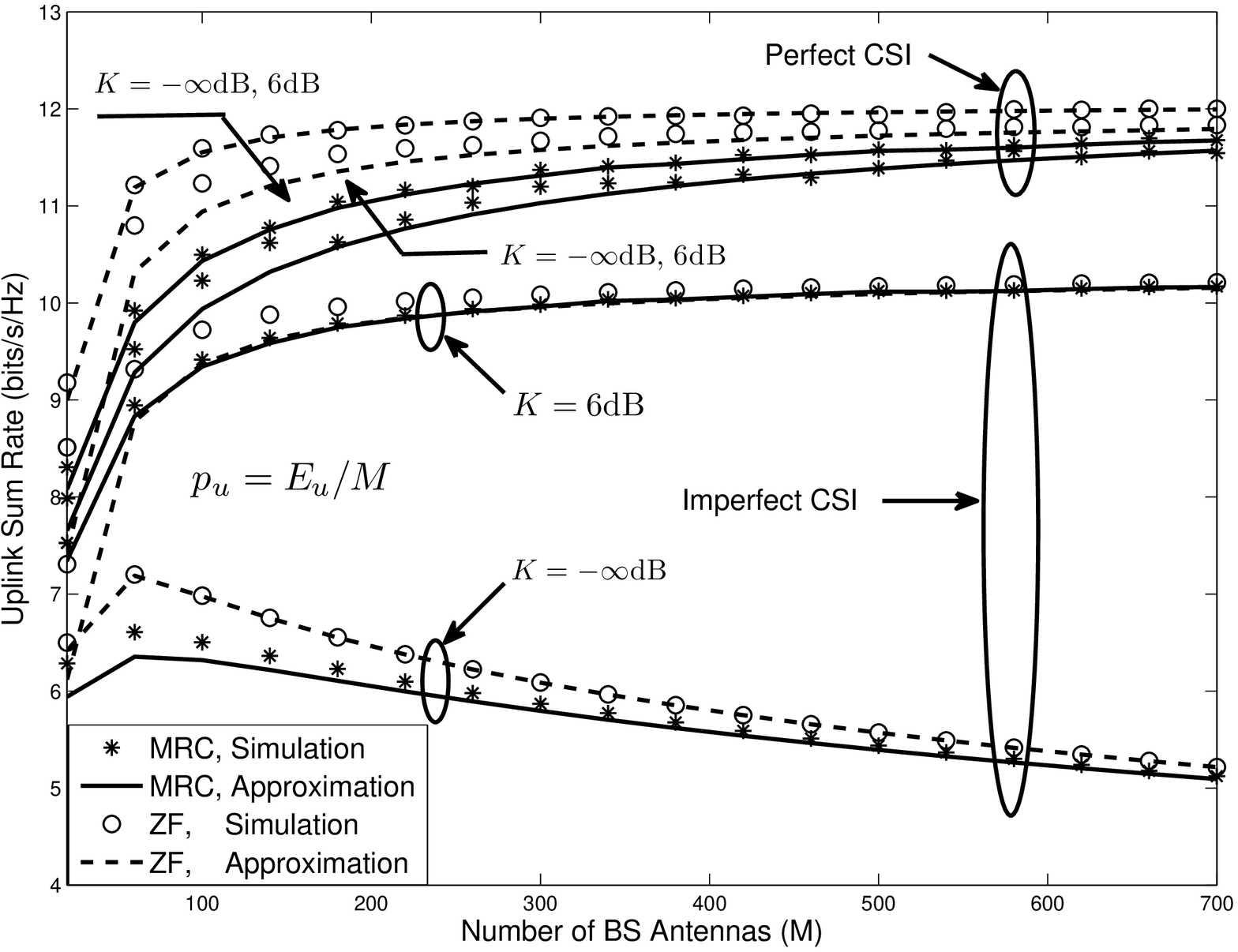}
\caption{Uplink sum rate per cell versus the number of BS antennas, with $N=10$ users and scaled-down power $p_u={E_u}/{M}$, where $E_u=20$dB.}\label{fig:fig 2}
\setlength{\abovecaptionskip}{0pt}
\setlength{\belowcaptionskip}{0pt}
\end{figure}

\begin{figure}
\centering
\includegraphics[scale=0.38]{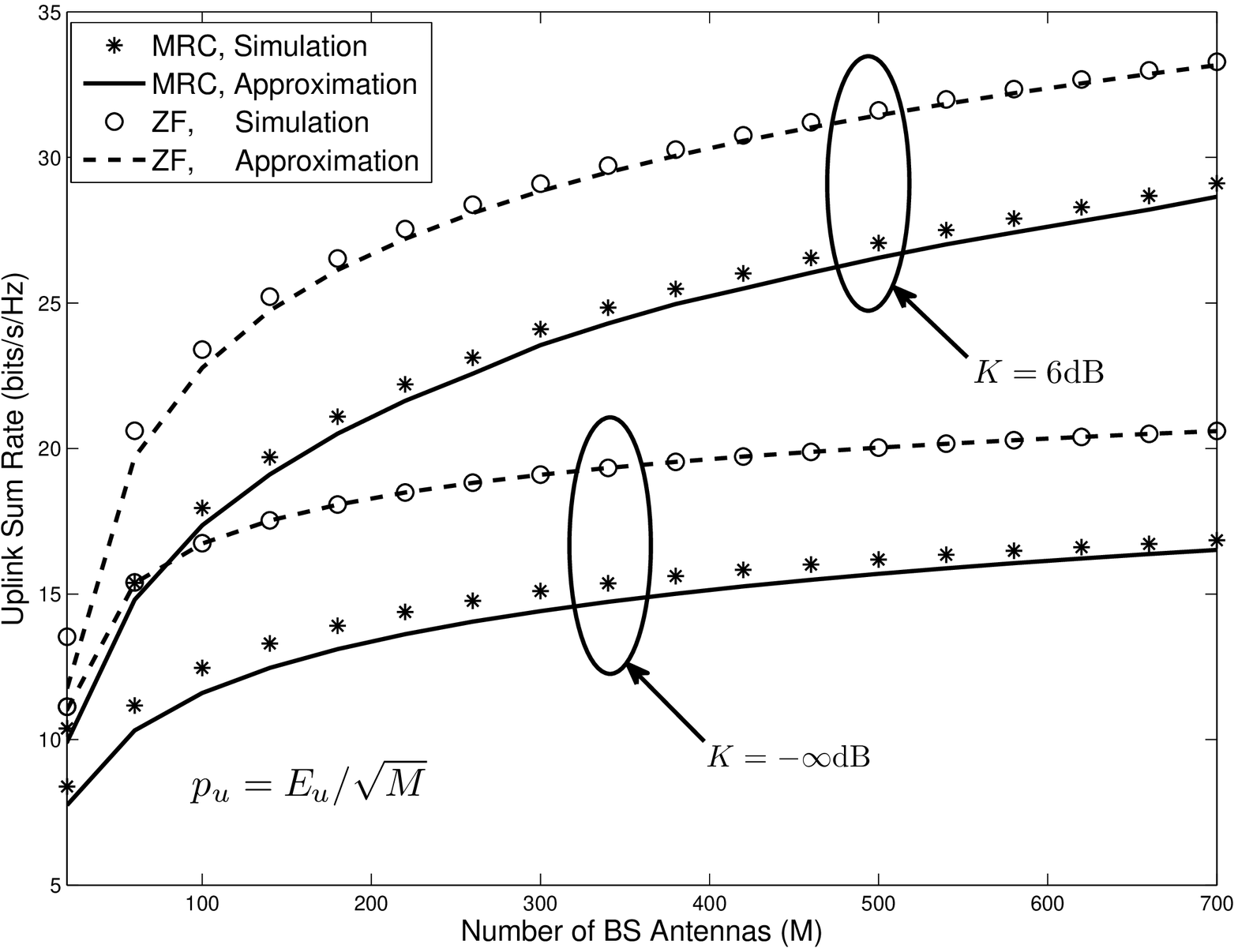}
\caption{Uplink sum rate per cell versus the number of BS antennas with $N=10$ users and scaled-down power $p_u={E_u}/{\sqrt M}$, where $E_u=20$dB.}\label{fig:fig 3}
\setlength{\abovecaptionskip}{0pt}
\setlength{\belowcaptionskip}{0pt}
\end{figure}

\begin{figure}
\centering
\includegraphics[scale=0.38]{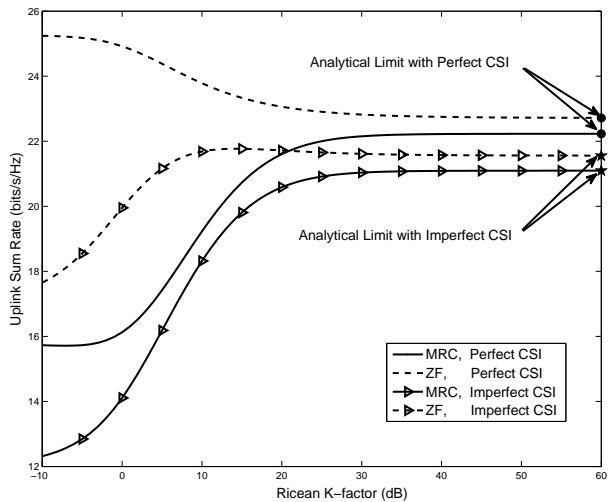}
\caption{Uplink sum rate per cell versus the Ricean $K$-factor with $N=10$ users, number of BS antennas $M=100$ and transmit power $p_u=10$dB.}\label{fig:fig 4}
\setlength{\abovecaptionskip}{0pt}
\setlength{\belowcaptionskip}{0pt}
\end{figure}

 For our simulations, we consider a cell with a radius of $1000$m and $N$ users distributed randomly and uniformly over the cell, with the exclusion of a central disk of radius $r_h=100$m. For convenience, we assume that every user has the same Ricean $K$-factor, denoted by $K$. The large-scale channel fading is modelled using $\beta_n={z_n}/{\left({r_n}/{r_h}\right)^v}$, where $z_n$ is a log-normal random variable with standard deviation $\sigma$, $v$ is the path loss exponent, and $r_n$ is the distance between the $n$th user and the BS. We have assumed that $\sigma=8$dB and $v=3.8$. Also, the transmitted data are modulated using orthogonal frequency-division multiplexing (OFDM). The parameters were chosen according to the LTE standard: an OFDM symbol interval of $T_s=500/7\approx 71.4\mu$s, a subcarrier spacing of $\Delta_f=15k$Hz, a useful symbol duration $T_u=1/\Delta_f \approx 66.7 \mu$s. We choose the channel coherence time to be $T_c=1m$s. As a result, the coherence time of the channel becomes $T=\frac{T_c}{T_s}\frac{T_u}{T_s-T_u}=196$ symbols.

In Fig.~\ref{fig:fig 1}, the simulated uplink sum rates per cell of MRC and ZF receivers in \eqref{mrc uplink rate}, \eqref{ZF uplink rate}, \eqref{MRC uplink rate} and \eqref{ZF imperfect uplink rate original} are compared with their corresponding analytical approximations in \eqref{MRC lowBound}, \eqref{ZF lowBound}, \eqref{MRC approximation} and \eqref{imperfect ZF lowBound}, respectively. Results are presented for two different scenarios -- perfect CSI and imperfect CSI. In this example, the number of users $N$ is assumed to be $10$ with the transmit power $p_u=10$dB. For each receiver, results are shown for different values of the Ricean $K$-factor, which are $0(-\infty$dB$)$, $3$dB, $6$dB, $10$dB, respectively. Clearly, in all cases, we see a precise agreement between the simulation results and our analytical results. For MRC receivers, as $K$ increases, the sum rate grows obviously for both perfect and imperfect CSI.
For ZF receivers, the sum rate has a noticeable growth only with imperfect CSI. We also find that both rates grow without bound, when no power normalization is being performed.

Fig.~\ref{fig:fig 2} investigates the power-scaling law when the transmit power of each user is scaled down by ${1}/{M}$. In the simulations, we choose $E_u=20$dB and $K$ has two different values, $0(-\infty$dB$)$ and $6$dB. With $p_u={E_u}/{M}$ and when $M \to\infty$, results show that the analytic approximations are the same as the exact results. In the case of perfect CSI, the sum rates of MRC and ZF receivers approach the same value, independent of the Ricean $K$-factor, as predicted in {Section \ref{subsec:perfect}}. For imperfect CSI, the sum rates of MRC and ZF receivers always tend to the same value as $M$ becomes large, which depends on $K$. {Note that when $K=0$, the sum rates do not have a monotonic behavior. This is because when $M$ is small, $p_u$ has not been cut down so much. Therefore, the improvement of the sum rate caused by the increased $M$ is greater than the loss of the sum rate caused by the reduced $p_u$. However, as $M$ gets larger, $p_u$ is cut down more aggressively. For the case $K=0$ with imperfect CSI, the decay of the sum rate brought by the reduced $p_u$ exceeds the growth of the sum rate brought by the increased $M$. Hence, the sum-rate curve will first rise and then drop as increasing $M$. In all other cases, with large $M$, there is a balance between the increase and decrease of the sum rate brought by the increased $M$ and scaling-down $p_u$, respectively. Therefore, these curves eventually saturate with an increased $M$.} Note that all these observations agree with \eqref{mrc imperfect k-ne-0 limit} and \eqref{zf imperfect k-ne-0 limit} and the curves for the case $K=0$ are consistent with the simulation results given by \cite[Fig. 2]{Ngo11}.

In Fig.~\ref{fig:fig 3}, we consider the same setting as in Fig.~\ref{fig:fig 2}, but the transmit power for each user is scaled down by a factor of ${1}/{\sqrt M}$. For convenience, we only show the results in the case with imperfect CSI. As can be seen, the sum rates of both MRC and ZF receivers converge to deterministic constants when $K=0$ but increase with $M$ when $K \ne 0$, which implies that the transmit power $p_u$ can be cut down more, just as predicted in {Section \ref{subsec:imperfect}}. Compared with Fig.~\ref{fig:fig 2}, we see that ZF performs much better than MRC. This is because ZF performs inherently better than MRC at high signal to noise ratio (SNR) and the transmit power here is only scaled down as $1/\sqrt M$, which makes the SNR relatively high. Again, all these observations are consistent with the conclusions in {Section \ref{subsec:imperfect}}. Interestingly, the convergence speed of the curves corresponding to the case $K=0$ is rather slow (approximately $\sqrt M$ times slower than that in the case considered in Fig.~\ref{fig:fig 2}).

To conclude this section, we study how the uplink sum rate varies with the Ricean $K$-factor in Fig.~\ref{fig:fig 4}. Due to the tightness between the simulated values and the approximations, here we only use the approximations in \eqref{MRC lowBound}, \eqref{ZF lowBound}, \eqref{MRC approximation} and \eqref{imperfect ZF lowBound} for analysis. The parameters were assumed to be $N=10$, $p_u=10$dB, and $M=100$. In all cases, as expected, the sum rates approximate the fixed value as $K \to \infty$. The two black dots represent the values obtained by \eqref{MRC K limitation} and \eqref{perfect ZF K}, respectively. The two black stars correspond to the values of the black dots multiplied by a coefficient $\left(T-\tau\right)/T$ for the calculation of the sum rate with imperfect CSI, as defined in \eqref{imperfect sum rate definition}. It is observed that every dot is indistinguishable from the curve limit, which not only means that our analytical results are accurate, but also that for both MRC and ZF receivers, the sum rates will tend to the same value as $K$ increases for both CSI cases. { We also find that the sum rates rise as increases with $K$, except for the case of ZF with perfect CSI. This is because the performance of ZF receivers can be severely limited if the channel matrix is ill-conditioned \cite{larsson09,artes03}. For the case of ZF receivers with perfect CSI, we experience no inter-user interference and no estimation-error. When $K$ grows, the channel matrix becomes identical to $\mathbf{\bar H}$, whose singular values have a large spread. Then, the condition number\footnote{Note that we are working with the 2-norm condition number, which is the ratio of the smallest to the largest singular value.} of $\mathbf{\bar H}$ attains very high values, which implies that is ill-conditioned. However, for the case of ZF with imperfect CSI, the main limiting factor is the estimation error. When $K$ grows large, channel estimation becomes far more robust, since quantities that were random before become deterministic. The same holds true for both cases of MRC receivers, where a higher Ricean $K$-factor reduces the effects of inter-user interference.}

\section{Conclusion}\label{sec:conclusion}
This paper worked out the achievable uplink rate of massive MIMO systems over Ricean fading channels with arbitrary-rank mean. We deduced new, tractable expressions for the uplink rate in the large-antenna limit along with exact approximations. Our analysis incorporated both ZF and MRC receivers and the cases of both perfect and imperfect CSI. These results were used to pursue a detailed analysis of the power-scaling law and of how the uplink rates change with the Ricean $K$-factor. We observed that with no degradation in the rate performance, the transmit power of each user can be at most cut down by a factor of ${1}/{M}$, except for the case of Rayleigh fading channels with imperfect CSI where the transmit power can only be scaled down up to a factor of ${1}/{\sqrt M}$. With scaling power and the same CSI quality, it was also shown that the uplink rates of MRC and ZF receivers will tend to the same constant value. Finally, with increasing Ricean $K$-factor and the same receiver, the uplink rates for perfect and imperfect CSI will also converge to the same fixed value.

\appendices
{\section{Proof of Lemma \ref{lemma 0}}\label{sec:proof of lemma 0}
 We know that
   \begin{small}
   \begin{equation}\label{approximation difference}
    \bbe \left\{\log_2\left(1+ \frac{X}{Y}\right)\right\}=\bbe\left\{\log_2\left(X+Y\right)\right\}-\bbe\left\{\log_2\left(Y\right)\right\}.
   \end{equation}
   \end{small}
  \hspace{-4pt} With the Jensen's inequality, we can get the following bounds:
   \begin{small}
    \begin{equation}\label{bounds signal 1}
    \log_2 \left(\frac{1}{\bbe\left\{\frac{1}{X+Y}\right\}}\right) \le \bbe\left\{\log_2\left(X+Y\right)\right\} \le \log_2\left(\bbe\left\{X+Y\right\}\right),
    \end{equation}
    \end{small}
    and
    \begin{small}
   \begin{equation}\label{bounds signal 2}
   \log_2 \left(\frac{1}{\bbe\left\{\frac{1}{Y}\right\}}\right) \le \bbe\left\{\log_2\left(Y\right)\right\} \le \log_2\left(\bbe\left\{Y\right\}\right).
   \end{equation}
   \end{small}
   \hspace{-4pt}Combing \eqref{bounds signal 1} and \eqref{bounds signal 2}, a lower and upper bound on \eqref{approximation difference} can be obtained as follows
   \begin{small}
\begin{multline}\label{approximation upper lower bound}
\log_2\left(\ntsp\frac{1}{\bbe\left\{\ntsp\frac{1}{X+Y}\ntsp\right\}}\ntsp\right)-\log_2\left(\bbe\left\{Y\right\}\right) \leq \bbe\left\{\log_2\left(\nmsp X\ntsp+\ntsp Y \nmsp \right)\right\}-\bbe\left\{\log_2\nmsp \left(Y \nmsp \right)\right\}\\ \leq \log_2\left(\bbe\left\{X+Y\right\}\right) -\log_2 \left(\frac{1}{\bbe\left\{\frac{1}{Y} \right\}}\right).
\end{multline}
\end{small}
\hspace{-4pt}Utilizing $\bbe\left\{\frac{1}{X+Y}\right\} \ge \frac{1}{\bbe\left\{X+Y\right\}}$ and $\bbe\left\{\frac{1}{Y}\right\} \ge \frac{1}{\bbe\left\{Y\right\}}$ and combining them with the left and right hand side of \eqref{approximation upper lower bound}, respectively, we get that
    \begin{small}
\begin{multline}\label{approximation upper lower bound 1}
\log_2\left(\ntsp\frac{1}{\bbe\left\{\ntsp\frac{1}{X+Y}\ntsp\right\}}\ntsp\right)-\log_2\left(\bbe\left\{Y\right\}\right) \leq \log_2\left(\bbe\ntsp \left\{\nmsp X\ntsp +\ntsp Y \nmsp \right\}\right) - \log_2 \left(\bbe \left\{Y\right\}\right)\\ \leq \log_2\left(\bbe\left\{ X + Y \right\}\right)  - \log_2\left(\frac{1}{\bbe\left\{\frac{1}{Y}\right\}}\right).
\end{multline}
\end{small}
    The above equation can be further written as
    \begin{small}
    \begin{multline}\label{approximation upper lower bound 2}
     \log_2\left(\frac{1}{\bbe\left\{\frac{1}{X+Y}\right\}}\right)-\log_2\left(\bbe\left\{Y\right\}\right) \leq \log_2\left(1+ \frac{\bbe \left\{ X   \right\}}{\bbe\left\{Y\right\}}\right)\\ \leq \log_2\left(\bbe\left\{ X + Y \right\}\right)  - \log_2\left(\frac{1}{\bbe\left\{\frac{1}{Y}\right\}}\right).
    \end{multline}
    \end{small}
    \hspace{-4pt}Comparing \eqref{approximation upper lower bound} and \eqref{approximation upper lower bound 2}, we find that $\log_2 \left(1+ \frac{\bbe\left\{X\right\}}{\bbe\left\{Y\right\}}\right)$ is a quantity that lies between the upper and lower bound of $\bbe \left\{\log_2\left(1+ \frac{X}{Y}\right)\right\}$. Therefore, it can be used as an approximation. \par

   Moreover, we can quantify the offset between these two bounds as
   \begin{small}
   \begin{equation}\label{bounds distance}
   d=\log_2 \left( {\bbe\left\{ {X + Y} \right\}\bbe\left\{ {\frac{1}{{X + Y}}} \right\}\bbe \left\{ Y \right\}\bbe \left\{ {\frac{1}{Y}} \right\}} \right).
   \end{equation}
   \end{small}
   \hspace{-4pt}By applying a Taylor series expansion of $\frac{1}{Y}$ around $\bbe\left\{Y\right\}$, we get that
   \begin{small}
\begin{multline}\label{taylor exapnsion}
\bbe\left\{Y\right\}\bbe\left\{\frac{1}{Y}\right\}\approx \bbe\left\{Y\right\} \\ \times \bbe\left\{\ntsp{\frac{1}{{\bbe\ntsp\left\{Y \right\}}}\ntsp - \ntsp \frac{1}{{{{ {\bbe^2\ntsp\left\{Y  \right\}} }}}}\left( {Y\ntsp  - \ntsp\bbe\ntsp \left\{ Y \right\}} \right) \ntsp +\ntsp \frac{1}{{{{{\bbe^3\ntsp \left\{Y\right\}} }}}}{{\left( {Y \ntsp -\ntsp \bbe\ntsp \left\{Y\right\}} \right)}^2}}\ntsp\right\}\\=1 + \frac{1}{\bbe^2\left\{Y\right\}}{\tt Var}\left\{Y\right\}
\end{multline}
\end{small}
   The substitution of \eqref{taylor exapnsion} into \eqref{bounds distance} yields
   \begin{small}
   \begin{multline}\label{bounds distance taylor}
   d \approx \log_2 \left( \left(1+ \frac{1}{\bbe^2\left\{X+Y\right\}}{\tt Var}\left\{X+Y\right\}\right)\right.\\
   \times \left.\left(1+ \frac{1}{\bbe^2\left\{Y\right\}}{\tt Var}\left\{Y\right\}\right)\right).
   \end{multline}
   \end{small}

   Since both $X$ and $Y$ are nonnegative, as $t_1$ and $t_2$ grow large, \begin{small}$\bbe\left\{X+Y\right\}$\end{small} and \begin{small}$\bbe\left\{Y\right\}$\end{small} will increase, and according to the law of large numbers, $X+Y$ and $Y$ will approach their means, that is,
    their variances will become small. Hence, the offset $d$ decreases with the increase of $t_1$ and $t_2$, that is, this approximation will be more and more accurate as the number of random variables of $X$ and $Y$ increase.}

\section{Proof of Lemma \ref{lemma 1}}\label{sec:proof of lemma 1}
According to the law of large numbers, we know that
\begin{equation}\label{H largeNumberLaw}
\frac{1}{M}\bh_n^H{\bh_i} - \frac{1}{M} \sum_{m = 1}^M \bbe\left\{{\left[ \bH \right]_{mn}^*{{\left[ \bH \right]}_{mi}}}\right\} \xrightarrow{a.s.} 0.
\end{equation}
The entries of $\bH$ can be obtained from \eqref{fast fading matrix} and \eqref{Hbar entries model} described in {Section \ref{sec:channel model}}. Hence,
\begin{small}
\begin{align}
\left[\bH\right]_{mn}
&=\sqrt {\frac{K_n}{{K_n + 1}}} {e^{ - j\left( {m - 1} \right)\pi \sin \left( {{\theta _n}} \right)}}+\sqrt {\frac{1}{{K_n + 1}}}\left( {{s_{mn}} + j{t_{mn}}} \right),\label{H entries}
\end{align}
\end{small}
\hspace{-5pt}with $s_{mn}$ and $t_{mn}$ representing the independent real and imaginary parts of $\left[{\bH_w}\right]_{mn}$, respectively, each with zero-mean and variance of ${1}/{2}$. After substituting \eqref{H entries} into \eqref{H largeNumberLaw} and with the definition
\begin{equation}\label{sigma s and t definition}
{e^{ - j\left( {m - 1} \right)\pi \sin \left( {{\theta _n}} \right)}}=\sigma_{mn},\quad{s_{mn}} + j{t_{mn}}=q_{mn},
\end{equation}
we have, as $M \to \infty$,
\begin{small}
\begin{align}
\frac{1}{M}\bh_n^H{\bh_i}
& - \frac{1}{M\sqrt{\left(K_n+1\right)\left(K_i+1\right)}}\sum\limits_{m = 1}^M \bbe\left\{ {\sqrt{K_n K_i}\sigma _{mn}^*{\sigma _{mi}} }\right.\notag\\
&\left.{+ \sqrt{K_n}\sigma _{mn}^*{q_{mi}}} { + \sqrt{K_i}{\sigma _{mi}}q_{mn}^* + q_{mn}^*{q_{mi}}} \right\}\xrightarrow{a.s.} 0.\label{H entries multiply}
\end{align}
\end{small}\par

If $i=n$, 
it can be easily obtained that $\frac{1}{M}\bh_n^H{\bh_n}\to 1$. If 
$i \ne n$, the last three terms of \eqref{H entries multiply} all become zero and the only remaining term is
\begin{small}
\begin{multline}\label{sum change}
\frac{\sqrt{K_n K_i}}{M\sqrt{\left(K_n+1\right)\left(K_i+1\right)}}\sum_{m = 1}^M{e^{j\left( {m - 1} \right)\pi\left[ {\sin\left({\theta _n}\right) - \sin\left({\theta _i}\right)} \right]}}  \\\mathop = \limits^{(a)}\frac{\sqrt{K_n K_i}}{M\sqrt{\left(K_n+1\right)\left(K_i+1\right)}}\phi_{ni}{e^{j\frac{\left({M - 1}\right)\pi}{2}\left[ {\sin\left({\theta _n}\right) - \sin\left({\theta _i}\right)} \right]}},
\end{multline}
\end{small}
\hspace{-10pt}where $\phi_{ni}$ is defined in \eqref{phi definition} and $(a)$ is obtained by using \cite[Eq. (14)]{ravindran07}. Due to the fact that ${{\sin \left( {\frac{M\pi}{2}\left[ {\sin\left({\theta _n}\right) - \sin\left({\theta _i}\right)} \right]} \right)}}{e^{j\frac{\left({M - 1}\right)\pi}{2}\left[ {\sin\left({\theta _n}\right) - \sin\left({\theta _i}\right)} \right]}}$ is bounded, we get as $M \to \infty$
\begin{small}
\begin{equation}\label{phi tend to 0}
\frac{\sin \left( {\frac{M\pi}{2}\left[ {\sin\left({\theta _n}\right) - \sin\left({\theta _i}\right)} \right]} \right)}{M}{e^{j\frac{\left({M - 1}\right)\pi}{2}\left[ {\sin\left({\theta _n}\right) - \sin\left({\theta _i}\right)} \right]}} \to 0.
\end{equation}
\end{small}
\hspace{-5pt}
Hence, we arrive at the desired result in \eqref{perfect H limit}.

\section{Proof of Lemma \ref{lemma 2}}\label{sec:proof of lemma 2}
Using \eqref{H entries multiply} in {Appendix \ref{sec:proof of lemma 1}}, we first consider the case $i=n$ and we have
\begin{align}\label{H entries multiply n}
\bh_n^H \bh_n &=\sum_{m = 1}^M {\left\{ {\frac{K_n}{{K_n + 1}} + \frac{{\sqrt {K_n} }}{{K_n + 1}}\sigma _{mn}^*{q_{mn}}{\rm{ }}} \right.}\notag\\
 &\left. { + \frac{{\sqrt {K_n} }}{{K_n + 1}}{\sigma _{mn}}q_{mn}^* + \frac{1}{{K_n + 1}}\left( {s_{mn}^2 + t_{mn}^2} \right)} \right\}.
\end{align}
Then, it is easily obtained that
\begin{equation}
\bbe\left\{\bh^H_n \bh_n\right\}=\sum_{m = 1}^M\left(\frac{K_n}{K_n+1}+\frac{1}{K_n+1}\right)=M.
\end{equation}

To evaluate the norm-square of a vector, we first should extract its real and imaginary parts. Note that $\bh_n^H\bh_n$ has only real parts. Hence, its norm-square can be obtained by squaring \eqref{H entries multiply n} directly as \eqref{H square n} (at the top of the next page).
\begin{figure*}[!t]
\begin{small}
\begin{align}
\left\|\bh_n\right\|^4&=\left|\bh^H_n \bh_n \right|^2
 =\frac{{4K_n}}{{{{\left( {K_n + 1} \right)}^2}}}{\left[ {\mathop \sum \limits_{m = 1}^M\left( {s_{mn}}\rho^c_{mn} - {t_{mn}}\rho^s_{mn}\right)}\right]^2}
+ \frac{1}{{{{\left( {K_n + 1} \right)}^2}}}{\left[ {\mathop \sum \limits_{m = 1}^M \left( {s_{mn}^2 + t_{mn}^2} \right)} \right]^2}
+ \frac{{2MK_n}}{{{{\left( {K_n + 1} \right)}^2}}}\mathop \sum \limits_{m = 1}^M \left( {s_{mn}^2 + t_{mn}^2} \right) \notag\\
&+ \frac{{4MK_n\sqrt {K_n} }}{{{{\left( {K_n + 1} \right)}^2}}}\mathop \sum \limits_{m = 1}^M \left( {{s_{mn}}\rho^c_{mn}} - {t_{mn}}\rho^s_{mn} \right)
+ \frac{{4\sqrt {K_n} }}{{{{\left( {K_n + 1} \right)}^2}}}\sum\limits_{{m_1} = 1}^M {\sum\limits_{{m_2} = 1}^M {\left( {{s_{m_1n}}\rho^c_{m_1 n} - {t_{m_1n}}\rho^s_{m_1n}} \right)} } \left( {s_{m_2n}^2 + t_{m_2n}^2} \right)+{\left( {\frac{{MK_n}}{{K_n + 1}}} \right)^2}.\label{H square n}
\end{align}
\end{small}
\hrulefill
\end{figure*}

Applying $\bbe\left\{\left(s_{mn}\right)^4\right\}=\bbe\left\{\left(t_{mn}\right)^4\right\}={3}/{4}$ in \eqref{H square n} and removing all the terms with zero mean, we can get
\begin{small}
\begin{align}\label{expectation of i=n}
\bbe\left\{\left\|\bh_n\right\|^4\right\}&=\left(\frac{{MK_n}}{{K_n + 1}}\right)^2\ntsp+\ntsp\frac{2MK_n}{\left(K_n+1\right)^2}\ntsp+\ntsp\frac{M^2+M}{\left(K_n+1\right)^2}\ntsp+\ntsp\frac{2M^2K_n}{\left(K_n+1\right)^2}\notag\\
&=\frac{2MK_n+M}{\left(K_n+1\right)^2}+M^2.
\end{align}
\end{small}
\hspace{-4pt}If $i \ne n$, from \eqref{H entries multiply} we know that
\begin{multline}\label{G multiply i'n}
\bh_n^H\bh_i=\frac{1}{\sqrt{(K_n+1)(K_i+1)}}\sum\limits_{m = 1}^M\left\{ {\sqrt{K_n K_i}\sigma _{mn}^*{\sigma _{mi}}}\right. \\
+ \left.{\sqrt{K_n}\sigma _{mn}^*{q_{mi}}}{ + \sqrt{K_i}{\sigma _{mi}}q_{mn}^* + q_{mn}^*{q_{mi}}}\right\}.
\end{multline}
To obtain the norm-square of \eqref{G multiply i'n}, both the real and imaginary parts should be extracted. Recalling \eqref{sum change}, we can rewrite \eqref{G multiply i'n} as
\begin{small}
\begin{align}
\left({\bh_n^H\bh_i}\right)_{\sf real}\ntsp&=\ntsp\frac{1}{\sqrt{(K_n+1)(K_i+1)}}\left\{\ntsp\sum\limits_{m = 1}^M {\ntsp\left[\ntsp \sqrt{K_n}\left({{s_{mi}}\rho^c_{mn} \ntsp- \ntsp {t_{mi}}\rho^s_{mn}}\right) \right.}\right. \notag\\
&+\left.{\left. \sqrt{K_i}\left({s_{mn}}\rho^c_{mi}- {t_{mn}}\rho^s_{mi}\right)+ \left( {{s_{mn}}{s_{mi}} + {t_{mn}}{t_{mi}}} \right)\right]}\right.\notag\\
&+\left.\sqrt{K_n K_i}\phi_{ni} \cos \left( {\frac{{M - 1}}{2}\pi \left[ {\sin \left( {{\theta _n}} \right) - \sin \left( {{\theta _i}} \right)} \right]} \right)\right\},\label{perfect real}\\
\left({\bh_n^H\bh_i}\right)_{\sf imag}\ntsp &=\ntsp\frac{1}{\sqrt{(K_n+1)(K_i+1)}}\left\{\ntsp\sum\limits_{m = 1}^M \ntsp\left[\sqrt{K_n}\left( {{t_{mi}}\rho^c_{mn} \ntsp+\ntsp {s_{mi}}\rho^s_{mn}}\right)\right.\right. \notag\\
&-\left. \sqrt{K_i}\left({s_{mn}}\rho^s_{mi} + {t_{mn}}\rho^c_{mi}\right)+{\left( {{s_{mn}}{t_{mi}} - {t_{mn}}{s_{mi}}} \right)}\right]\notag\\
&+\left.\sqrt{K_n K_i}\phi_{ni} \sin \left( {\frac{{M - 1}}{2}\pi \left[ {\sin \left( {{\theta _n}} \right) - \sin \left( {{\theta _i}} \right)} \right]} \right)\right\}\label{perfect imag}.
\end{align}
\end{small}
\hspace{-5pt}Then, with
\begin{equation}\label{expectation of i'n}
\bbe\left\{\left|\bh_n^H\bh_i\right|^2\right\}=\bbe\left\{\left({\bh_n^H\bh_i}\right)_{\sf real}^2+\left({\bh_n^H\bh_i}\right)_{\sf imag}^2\right\},
\end{equation}
substituting \eqref{perfect real} and \eqref{perfect imag} into \eqref{expectation of i'n} and removing the terms with zero expectation, we get the final result as
\begin{equation}
\bbe\left\{\left|\bh_n^H\bh_i\right|^2\right\}=\frac{{{K_n K_i}{\phi ^2_{ni}} + M\left(K_n+K_i\right) + M}}{{{{\left( {K_n + 1} \right)\left(K_i+1\right)}}}}.
\end{equation}
As we have obtained all expectations used in {\it Lemma \ref{lemma 2}}, we now conclude the proof.

\section{Proof of Lemma \ref{lemma 3}}\label{sec:proof of lemma 3}
Following the similar procedure as in the proof of {\it Lemma \ref{lemma 1}}, we start by giving the entries of the channel matrix. The imperfect CSI model has been described in \eqref{Imperfect CSI model} of {Section \ref{sec:imperfect CSI model}}, which yields
\begin{equation}
{\left[ \mathbf{\hat G} \right]_{mn}} = \sqrt {\frac{K_n}{{K_n + 1}}} {\left[ \mathbf{\bar G} \right]_{mn}} + \sqrt {\frac{1}{{K_n + 1}}} {\left[ {{{\mathbf{\hat G}}_w}} \right]_{mn}}.
\end{equation}
The first term which does not need to be estimated is the same as in the case with perfect CSI. For $\mathbf{\hat G}_w$, we use the MMSE estimation to get it and the details of the method have been introduced in {Section \ref{sec:imperfect CSI model}}. Therefore, we have
\begin{align}\label{imperfect G entries}
{\left[ \mathbf{\hat G} \right]_{mn}}&=\sqrt {\frac{K_n}{{K_n + 1}}} \sqrt {{\beta _n}} {\sigma _{mn}} \notag\\
&+ \sqrt {\frac{1}{{K_n + 1}}} {\eta _n}\left[ {\sqrt {{\beta _n}}  {q_{mn}} + \frac{1}{{\sqrt {{p_p}} }}\left( {s{w_{mn}} + jt{w_{mn}}} \right)} \right]
\end{align}
where $\sigma_{mn}$, $s_{mn}$ and $t_{mn}$ have all been defined in \eqref{sigma s and t definition} of {Appendix \ref{sec:proof of lemma 2}}, $sw_{mn}$ and $tw_{mn}$ represent the independent real and imaginary part of $\left[\mathbf{W}\right]_{mn}$ with zero-mean and variance ${1}/{2}$, and $\eta_n$ is defined in \eqref{eta definition}.
By the law of large numbers, it can be got that
\begin{equation}\label{law of large numbers imperfect i=n}
\frac{1}{M}\bhg_n^H{{\bhg}_n}- \frac{1}{M}\sum\limits_{m = 1}^M \bbe{\left\{ {\left[ \bhG \right]_{mn}^*{{\left[ \bhG \right]}_{mn}}} \right\}} \xrightarrow{a.s.} 0.
\end{equation}
With \eqref{law of large numbers imperfect i=n} and the definition
\begin{equation}
sw_{mn}+jtw_{mn}=qw_{mn},
\end{equation}
we can obtain \eqref{imperfect G multiply n} (at the top of the next page), as $M \to \infty$,
\begin{figure*}[!t]
\begin{small}
\begin{align}
\frac{1}{M}\bhg_n^H{{\bhg}_n}
&-\frac{1}{M}\sum\limits_{m = 1}^M \bbe\left\{\frac{K_n}{{K_n + 1}}{\beta _n} + \frac{{\sqrt {K_n} }}{{K_n + 1}}\sqrt {{\beta _n}} {\eta _n}\sigma _{mn}^*\left( {\sqrt {{\beta _n}} {q_{mn}} + \frac{1}{{\sqrt {{p_p}} }}q{w_{mn}}} \right)+\frac{1}{{K_n + 1}}\frac{\sqrt{\eta_n^2\beta_n}}{\sqrt{p_p}}\left( {q_{mn}^*q{w_{mn}} + {q_{mn}}qw_{mn}^*} \right)\right.\notag\\
&+\frac{{\sqrt {K_n} }}{{K_n + 1}}\sqrt {{\beta _n}} {\eta _n}{\sigma _{mn}}\left( {\sqrt {{\beta _n}} q_{mn}^* + \frac{1}{{\sqrt {{p_p}} }}qw_{mn}^*} \right)
\left.+\frac{1}{{K_n + 1}}\eta _n^2\left[ {{\beta _n}\left( {s_{mn}^2 + t_{mn}^2} \right)}
+\frac{1}{p_p}\left(sw_{mn}^2+tw_{mn}^2\right)\right]\right\}\xrightarrow{a.s.} 0.\label{imperfect G multiply n}
\end{align}
\end{small}
\hrulefill
\end{figure*}
Evaluating the expectation of all terms in \eqref{imperfect G multiply n} and removing the terms with zero expectation, \eqref{imperfect G multiply n} can be simplified as
\begin{multline}\label{proof of lemma 2 result 1}
\frac{1}{M}\bhg_n^H{{\bhg}_n}-\\ \frac{1}{M}\sum_{m = 1}^M\left[\frac{K_n}{K_n+1}\beta_n+\frac{1}{K_n+1}\eta_n^2\left(\beta_n+\frac{1}{p_p}\right)\right]\xrightarrow{a.s.}0,
\end{multline}
which yields that
\begin{equation}
\frac{1}{M}\bhg_n^H{{\bhg}_n} \xrightarrow{a.s.} \frac{\beta_n}{K_n+1}\left(K_n+\eta_n\right).
\end{equation}

Now consider $\bhg_n^H\bhg_i$. Following the same operations as in \eqref{imperfect G multiply n}, with
\begin{equation}
\frac{1}{M}\bhg_n^H{{\bhg}_i} - \frac{1}{M}\sum\limits_{m = 1}^M\bbe{\left\{ {\left[ \bhG \right]_{mn}^*{{\left[ \bhG \right]}_{mi}}} \right\}}\xrightarrow{a.s.} 0
\end{equation}
we find \eqref{imperfect G multiply i} (at the top of the next page).
\begin{figure*}[!t]
\begin{small}
\begin{align}
\frac{1}{M}\bhg_n^H{{\bhg}_i}
&-\frac{1}{M\sqrt{(K_n+1)(K_i+1)}}\sum\limits_{m = 1}^M\bbe\left\{\sqrt{K_nK_i}\sqrt {{\beta _n}{\beta _i}} \sigma _{mn}^*{\sigma _{mn}}+ \sqrt{K_n}\sqrt {{\beta _n}} {\eta _i}\sigma _{mn}^*\left( {\sqrt {{\beta _i}} {q_{mi}} + \frac{1}{{\sqrt {{p_p}} }}q{w_{mi}}} \right)\right.\notag\\
&\left.+\sqrt{K_i}\sqrt {{\beta _i}} {\eta _n}{\sigma _{mi}}\left( {\sqrt {{\beta _n}} q_{mn}^* + \frac{1}{{\sqrt {{p_p}} }}qw_{mn}^*} \right)+{\eta _n}{\eta _i}\left( {q_{mn}^*{q_{mi}} + qw_{mn}^*q{w_{mi}} + q_{mn}^*q{w_{mi}} + {q_{mi}}qw_{mn}^*} \right)\right\} \xrightarrow{a.s.} 0.\label{imperfect G multiply i}
\end{align}
\end{small}
\hrulefill
\end{figure*}
Since the entries of $\bH$ and $\mathbf W$ are all i.i.d.~zero-mean with unit-variance, many terms in \eqref{imperfect G multiply i} have zero expectation. The only remaining term is the element of the channel mean matrix. Hence, \eqref{imperfect G multiply i} can be written as
\begin{multline}
\frac{1}{M}\bhg_n^H{{\bhg}_i} -\\ \frac{1}{M}\sum\limits_{m = 1}^M\frac{\sqrt{K_nK_i}}{{\sqrt{(K_n + 1)(K_i+1)}}}\sqrt {{\beta _n}{\beta _i}} \sigma _{mn}^*{\sigma _{mi}}\xrightarrow{a.s.} 0,
\end{multline}
which can be simplified as
\begin{multline}\label{proof of lemma 3 imperfect g}
\frac{1}{M}\bhg_n^H{{\bhg}_i} -\\\frac{\sqrt {{\beta _n}{\beta _i}}\sqrt{K_nK_i}\phi_{ni}}{M\sqrt{(K_n+1)(K_i+1)}}e^{\frac{M-1}{2}\pi\left[\sin\left(\theta_n\right)-\sin\left(\theta_i\right)\right]} \xrightarrow{a.s.} 0.
\end{multline}
According to \eqref{phi tend to 0}, we know that, as $M \to \infty$
\begin{equation}
\frac{\phi_{ni}}{M}e^{\frac{M-1}{2}\pi\left[\sin\left(\theta_n\right)-\sin\left(\theta_i\right)\right]} \to 0.
\end{equation}
As a consequence, \eqref{proof of lemma 3 imperfect g} can be further simplified as
\begin{equation}
\frac{1}{M}\bhg_n^H\bhg_i \xrightarrow{a.s.} 0.
\end{equation}

\section{Proof of Lemma \ref{lemma 4}}\label{sec:proof of lemma 4}
The expectation of $\bhg_n^H\bhg_n$ can be easily obtained by \eqref{proof of lemma 2 result 1} in {Appendix \ref{sec:proof of lemma 3}}. Hence, we have
\begin{equation}
\bbe\left\{\bhg_n^H\bhg_n\right\}=\frac{M\beta_n}{K_n+1}\left(K_n+\eta_n\right).
\end{equation}
Following the same procedure as in the proof of {\it Lemma \ref{lemma 2}}, we can obtain the remaining three norm-square expectations. First consider the case $i=n$. According to \eqref{imperfect G multiply n}, we have \eqref{imperfect expection in} (at the top of the next page).
\begin{figure*}[!t]
\begin{small}
\begin{align}
\bbe\left\{\left|\bhg_n^H\bhg_n\right|^2\right\}&=\bbe\left\{\sum\limits_{m = 1}^M\frac{K_n}{{K_n + 1}}{\beta _n} + \frac{{\sqrt {K_n} }}{{K_n + 1}}\sqrt {{\beta _n}} {\eta _n}\sigma _{mn}^*\left[ {\sqrt {{\beta _n}} {q_{mn}} + \frac{1}{{\sqrt {{p_p}} }}q{w_{mn}}} \right]+\frac{1}{{K_n + 1}}\eta _n^2\left( {q_{mn}^*q{w_{mn}} + {q_{mn}}qw_{mn}^*} \right)\right.\notag\\
&\left.+\frac{{\sqrt {K_n} }}{{K_n + 1}}\sqrt {{\beta _n}} {\eta _n}{\sigma _{mn}}\left[ {\sqrt {{\beta _n}} q_{mn}^* + \frac{1}{{\sqrt {{p_p}} }}qw_{mn}^*} \right]
+\frac{1}{{K_n + 1}}\eta _n^2\left[ {{\beta _n}\left( {s_{mn}^2 + t_{mn}^2} \right)} +\frac{1}{p_p}\left(sw_{mn}^2+tw_{mn}^2\right)\right]\right\}^2.\label{imperfect expection in}
\end{align}
\end{small}
\hrulefill
\end{figure*}
After expanding the above equation and removing all the terms with zero expectation, the remaining terms are written as \eqref{imperfect expection in result} (at the top of the next page).
\begin{figure*}[!t]
\begin{small}
\begin{align}
\bbe\left\{\left|\bhg_n^H\bhg_n\right|^2\right\}&={\left( {\frac{{MK_n}}{{K_n + 1}}} \right)^2}\beta _n^2 + \frac{{2MK_n}}{{{{\left( {K_n + 1} \right)}^2}}}{\beta _n}\eta _n^2\left( {{\beta _n} + \frac{1}{{{p_p}}}} \right) + \frac{{2M{\beta _n}}}{{{{\left( {K_n + 1} \right)}^2}}}\eta _n^4+ \left( {{M^2} + M} \right)\left( {\beta _n^2 + \frac{1}{{p_p^2}}} \right)\frac{{\eta _n^4}}{{{{\left( {K_n + 1} \right)}^2}}}\notag\\
 &+ \frac{{2{M^2}K_n\eta _n^2}}{{{{\left( {K_n + 1} \right)}^2}}}\left( {\beta _n^2 + \frac{{{\beta _n}}}{{{p_p}}}} \right) + \frac{{2{M^2}\eta _n^4}}{{{{\left( {K_n + 1} \right)}^2}}}\frac{{{\beta _n}}}{{{p_p}}}
=\frac{{\beta _n^2}}{{{{\left( {K_n + 1} \right)}^2}}}\left[ {{M^2}{K_n^2} + \left( {2MK_n + 2M^2{K_n}} \right){\eta _n} + \left( {{M^2} + M} \right)\eta _n^2} \right].\label{imperfect expection in result}
\end{align}
\end{small}
\hrulefill
\end{figure*}

For the case $i \ne n$, we use the same method as in the proof of {\it Lemma \ref{lemma 2}}. With \eqref{imperfect G multiply i}, it is noted that the inner product $\bhg_n^H\bhg_i$ has non-zero real and imaginary parts.
The result then follows trivially by some algebraic manipulations.

\begin{biography}[{\includegraphics[width=1in,height=1.25in,clip,keepaspectratio]{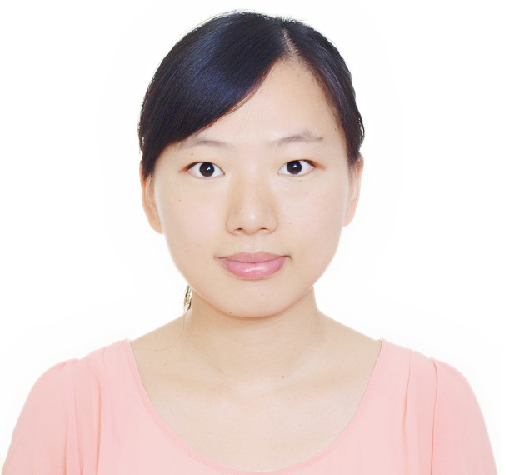}}]{Qi Zhang}
 received the B.S. degree in Electrical \& Information Engineering from Nanjing University of Posts \& Telecommunications, Nanjing, China, in 2010. She is currently working toward the Ph.D. degree in Communication \&
Information System at the Nanjing University of Posts \& Telecommunications, China. Her research
interests include massive MIMO systems and space-time wireless communications.
\end{biography}

\begin{biography}[{\includegraphics[width=1in,height=1.25in,clip,keepaspectratio]{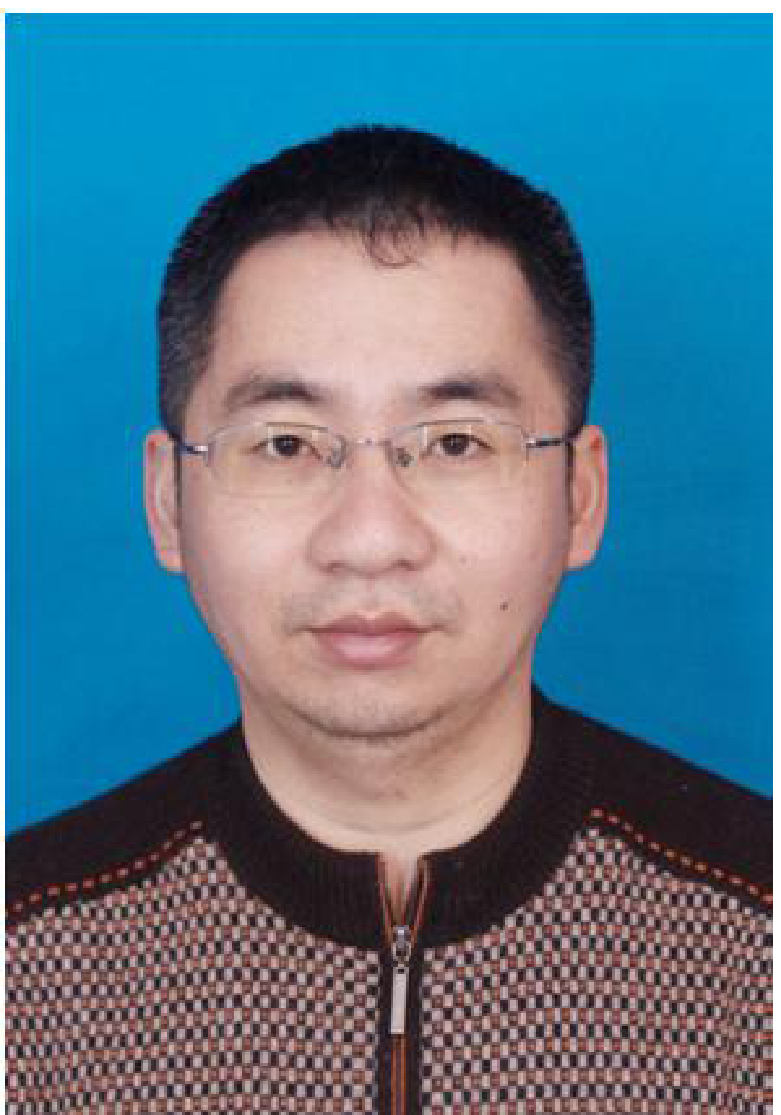}}]{Shi Jin}
(S'06-M'07) received the B.S. degree in communications engineering from Guilin University of Electronic Technology, Guilin, China, in 1996, the M.S. degree from Nanjing University of Posts and Telecommunications, Nanjing, China, in 2003, and the Ph.D. degree in communications and information systems from the Southeast University, Nanjing, in 2007. From June 2007 to October 2009, he was a Research Fellow with the Adastral Park Research Campus, University College London, London, U.K. He is currently with the faculty of the National Mobile Communications Research Laboratory, Southeast University. His research interests include space time wireless communications, random matrix theory, and information theory. He serves as an Associate Editor for the IEEE Transactions on Wireless Communications, and IEEE Communications Letters, and IET Communications. Dr. Jin and his co-authors have been awarded the 2011 IEEE Communications Society Stephen O. Rice Prize Paper Award in the field of communication theory and a 2010 Young Author Best Paper Award by the IEEE Signal Processing Society.
\end{biography}

\begin{biography}[{\includegraphics[width=1in,height=1.25in,clip,keepaspectratio]{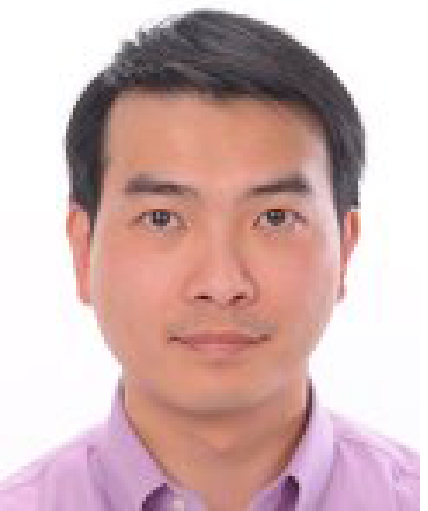}}]{Kai-Kit Wong}
(S'99-M'01-SM'08) received the BEng, the MPhil, and the PhD degrees, all in Electrical and Electronic Engineering, from the Hong Kong University of Science \& Technology, Hong Kong, in 1996, 1998, and 2001, respectively. Since August 2006, he has been with University College London, first at Adastral Park Campus and at present the Department of Electronic \& Electrical Engineering, where he is a Reader in Wireless Communications.

Dr Wong is a Senior Member of IEEE and Fellow of the IET. He is on the editorial board of IEEE Wireless Communications Letters, IEEE Communications Letters, IEEE ComSoc/KICS Journal of Communications and Networks, and IET Communications. He also previously served as Editor for IEEE Transactions on Wireless Communications from 2005-2011 and IEEE Signal Processing Letters from 2009-2012.
\end{biography}

\begin{biography}[{\includegraphics[width=1in,height=1.25in,clip,keepaspectratio]{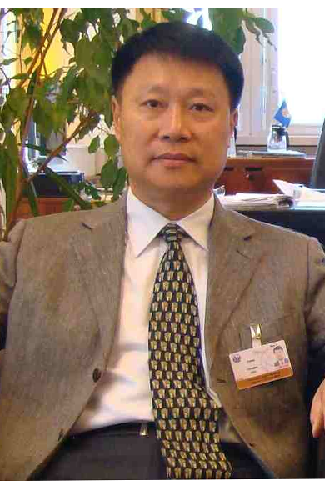}}]{Hongbo Zhu}
received the bachelor degree in Telecommunications Engineering from the Nanjing University of Posts \& Telecommunications, Nanjing, China and Ph.D. degree in Information \& Communications Engineering from Beijing University of Posts \& Telecommunications, Beijing, China, in 1982 and 1996, respectively. He is presently working as a Professor and Vice-president in Nanjing University of Posts \& Telecommunications, Nanjing, China. He is also the head of the Coordination Innovative Center of IoT Technology and Application (Jiangsu), which is the first governmental authorized Coordination Innovative Center of IoT in China. He also serves as referee or expert in multiple national organizations and committees.

He has published more than 200 papers on information and communication area, such as IEEE Trans. Presently, he is leading a big group and multiple funds on IoT and wireless communications with current focus on architecture and enabling technologies for Internet of Things.
\end{biography}

\begin{biography}[{\includegraphics[width=1in,height=1.25in,clip,keepaspectratio]{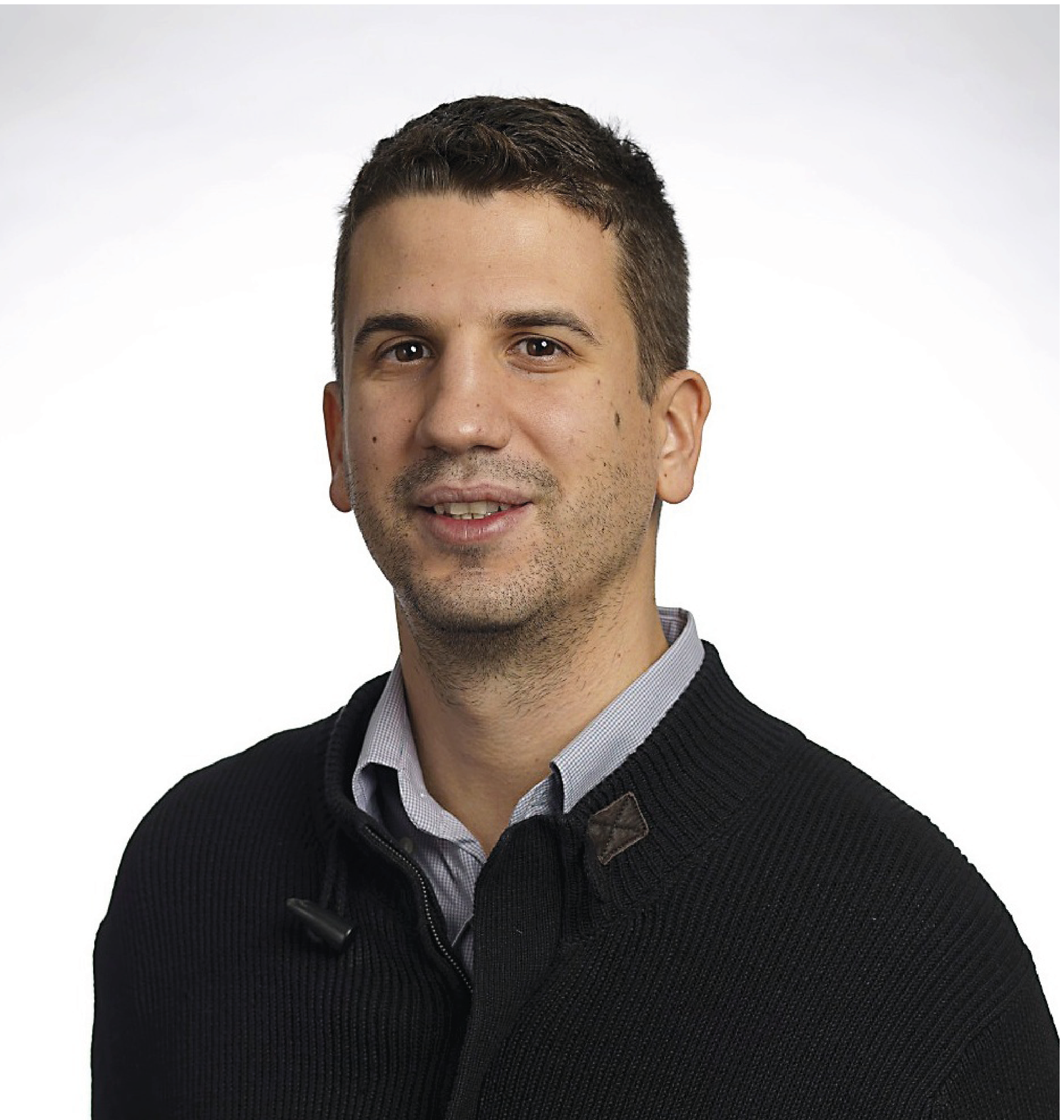}}]{Michail Matthaiou}
(S'05--M'08--SM'13) was born in Thessaloniki, Greece in 1981. He obtained the Diploma degree (5 years) in Electrical and Computer Engineering from the Aristotle University of Thessaloniki, Greece in 2004. He then received the M.Sc. (with distinction) in Communication Systems and Signal Processing from the University of Bristol, U.K. and Ph.D. degrees from the University of Edinburgh, U.K. in 2005 and 2008, respectively. From September 2008 through May 2010, he was with the Institute for Circuit Theory and Signal Processing, Munich University of Technology (TUM), Germany working as a Postdoctoral Research Associate.
He is currently a Senior Lecturer at Queen's University Belfast, U.K. and also holds an adjunct Assistant Professor position
at Chalmers University of Technology, Sweden. His research interests span signal processing for wireless communications, massive MIMO, hardware-constrained communications, and performance analysis of fading channels.

Dr. Matthaiou is the recipient of the 2011 IEEE ComSoc Young Researcher Award for the Europe, Middle East and Africa Region
and a co-recipient of the 2006 IEEE Communications Chapter Project Prize for the best M.Sc. dissertation in the area
of communications. He was an Exemplary Reviewer for \textsc{IEEE Communications Letters} for 2010. He has been a member of
Technical Program Committees for several IEEE conferences such as ICC, GLOBECOM, VTC etc. He currently serves as an Associate Editor for
the \textsc{IEEE Transactions on Communications}, \textsc{IEEE Communications Letters} and was the Lead Guest Editor of the special issue on ``Large-scale multiple antenna wireless systems'' of the \textsc{IEEE Journal on Selected Areas in Communications}. He is an associate member of the IEEE Signal Processing Society SPCOM and SAM technical committees.
\end{biography}


\end{document}